\documentclass[10pt,aps,prd,twocolumn,showpacs,superscriptaddress,eqsecnum,longbibliography,nofootinbib]{revtex4-2}
\usepackage{mathrsfs,amsmath,amsthm,latexsym,amssymb,amsfonts,epsfig,cancel,enumerate,graphicx,subcaption,txfonts,overpic}
\usepackage{multirow}
\usepackage{xcolor}
\definecolor{navy}{RGB}{0,0,150}
\usepackage[colorlinks, linkcolor=blue, citecolor=blue, urlcolor=blue, plainpages=false, pdfstartview=FitH]{hyperref}
\usepackage{appendix}
\allowdisplaybreaks
\newcommand{\GZU}{College of Physics, Guizhou University, Guiyang 550025, China}
\newcommand{\HU}{College of Physics and Science and Technology, Hebei University, Baoding 071002, China}
\newcommand{\SU}{School of Physics and Electronic Engineering, Shanxi Normal University, Taiyuan 030031, China}
\begin{document}
	
	\title{Rotating Charged Black Holes with Scalar Hair Constructed via the Newman-Janis Algorithm: Accretion Disk Structure and Shadow Characteristics}
	
	\author{Ziqiang Cai}
	\email{gs.zqcai24@gzu.edu.cn}
	\affiliation{\GZU}
	
	\author{Zhenglong Ban}
	\email{zlban123@163.com}
	\affiliation{\HU}
	
	\author{Qi-Qi Liang}
	\email{13985671614@163.com}
	\affiliation{\GZU}
	
	\author{Haiyuan Feng}
	\thanks{Corresponding author}
	\email{fenghaiyuanphysics@gmail.com}
	\affiliation{\SU}
	
	\author{Zheng-Wen Long}
	\thanks{Corresponding author}
	\email{zwlong@gzu.edu.cn}
	\affiliation{\GZU}
	
	\begin{abstract}
		In this paper, we generate a rotating charged black hole (BH) with scalar hair via the Newman–Janis algorithm (NJA) and study its thin accretion disk and shadow. The structure of the event horizon and ergosurface is analyzed in detail, revealing how the charge parameter $Q$ and scalar hair parameter $s$ influence the spacetime geometry. We analyze the energy flux and temperature distribution of the accretion disk, finding that increasing either $Q$ or $s$ leads to higher energy flux and peak temperature. The BH shadow is also examined, showing that its apparent size decreases monotonically with increasing 
		$Q$ or $s$. Notably, in the near-extremal regime, the shadow develops a distinctive cuspy edge, indicative of strong light bending in the scalarized and charged spacetime. By comparing the theoretically predicted shadow diameter with Event Horizon Telescope (EHT) observations of Sgr A$^{*}$, we derive observational constraints on the model parameters. For inclination angles of $17^{\circ}$ and $90^{\circ}$, a joint analysis constrains the charge parameter to $0<Q<0.522745$ (at fixed $s=0.1$) and the scalar hair parameter to $0<s<0.283373$ (at fixed $Q=0.3$). Our results demonstrate how scalar hair and electric charge leave imprints on accretion disk emissions and black hole shadows, offering new observational signatures for testing gravity theories beyond general relativity.
	\end{abstract}
	%\keywords{}
	
	\maketitle
	
	\section{Introduction}
	
	The EHT Collaboration has achieved monumental milestones in astrophysics with the first images of supermassive BHs. In April 2019, the EHT released the shadow image of M87$^{*}$, providing the first direct visual evidence of a BH's existence \cite{EventHorizonTelescope:2019dse,EventHorizonTelescope:2019uob,EventHorizonTelescope:2019jan,EventHorizonTelescope:2019ths,EventHorizonTelescope:2019pgp,EventHorizonTelescope:2019ggy} and confirming their presence in extreme environments. This success was followed by the release of an image of Sgr A$^{*}$ \cite{EventHorizonTelescope:2022wkp}, the supermassive BH at the center of our galaxy, in May 2022. Building on these achievements, the EHT has also released polarized images of M87$^{*}$, further enhancing our understanding of BHs and their surroundings. These observations not only represent significant advancements in observational astronomy but also offer critical tests for theories of gravity under extreme conditions. The physical processes governing BH accretion, a key mechanism driving their observable properties, are intimately linked to the structure of accretion disks around them.
	
	Compact objects such as BHs accumulate mass through accretion, which is facilitated by the presence of an accretion disk. An accretion disk consists of diffuse material like gas and dust that spirals inward toward the central object under the influence of gravitational forces. This process releases gravitational energy in the form of radiation. The foundational model for geometrically thin and optically thick accretion disks was first established by Shakura and Sunyaev. Subsequently, this model was extended to incorporate the effects of general relativity (GR) through the contributions of Page, Novikov, and Thorne \cite{Shakura:1972te,Page:1974he}. A considerable body of research has explored the physical attributes of matter forming thin accretion disks across different spacetime backgrounds \cite{Harko:2009rp,Chen:2011wb,Liu:2021yev,Heydari-Fard:2021ljh,Karimov:2018whx,Chen:2011rx,Kazempour:2022asl,Gyulchev:2019tvk,Wu:2024sng,Feng:2024iqj,Liu:2024brf}. The distinctive features present in the energy flux and the emitted spectrum from these disks provide crucial information about BHs and can also be used to test the validity of modified gravity theories \cite{Bambi:2015kza}. Complementary to these insights from accretion disk radiation, the geometric signature of BHs—their shadow—offers a unique probe of spacetime geometry under the strongest gravitational fields.
	
	The BH shadow, a two-dimensional dark area observed in the celestial sphere, arises from the strong gravitational influence of the BH. Its shape and dimensions are largely dictated by the specific geometry of the BH spacetime. Thus, strong lensing images and shadows provide an exciting opportunity to both investigate the nature of compact objects and test whether the gravitational field around them conforms to the Schwarzschild or Kerr geometry. The theoretical study of black hole shadows originated with Synge \cite{Synge:1966okc}, who calculated the shadow of a Schwarzschild black hole, and was advanced by Luminet \cite{Luminet:1979nyg} through detailed ray-tracing analyses. Bardeen \cite{Bardeen:1973tla} later pioneered the investigation of shadows for Kerr black holes, revealing their distinctive dependence on spin and observer inclination. Subsequently, shadows have been established as a powerful probe of strong-field gravity, encoding information about the black hole's mass, angular momentum, charge, inclination, and accretion flow structure. Furthermore, they provide a unique testbed for fundamental physics, enabling tests of general relativity and constraints on dark matter, cosmic acceleration, and potential extra dimensions. For further examples, see \cite{Stuchlik:2019uvf,Yumoto:2012kz,Abdujabbarov:2016hnw,Amir:2016cen,Sharif:2016znp,Wei:2013kza,Abdujabbarov:2012bn,Neves:2020doc,Amarilla:2013sj,Atamurotov:2013sca,Mishra:2019trb,Papnoi:2014aaa,Abdujabbarov:2015rqa,Kumar:2020owy,Liu:2020ola,Sarikulov:2022atq,Ban:2024qsa,Yang:2024nin,Wang:2025ihg,Zeng:2025kqw,Yunusov:2024xzu,Zahid:2024hwi,Li:2024abk,Chen:2023wzv,Liu:2024lve,Zahid:2025cfu,Lambiase:2024lvo,Ahmed:2025zdc,Zheng:2024ftk,Siahaan:2025nlq}.
	
	Deriving exact rotating BH solutions from the Einstein field equations is highly challenging due to the complexity of the resulting nonlinear, coupled partial differential equations. The NJA offers a valuable alternative, enabling the construction of stationary, axisymmetric rotating BH spacetimes from static, spherically symmetric seed metrics \cite{Newman:1965tw}. Compared to static counterparts, rotating BHs introduce frame-dragging and modify spacetime curvature, profoundly affecting both photon trajectories (shaping shadows) and energy distributions in accretion disks.
	
	According to the no-hair theorem, an asymptotically flat BH in classical gravity is fully characterized by only three parameters: its mass $M$, electric charge $Q$, and angular momentum $J$ \cite{Ruffini:1971bza}. The inclusion of scalar fields offers a powerful probe of the no-hair conjecture. In minimally coupled theories, consistency with asymptotic and horizon boundary conditions generally excludes stable scalar hair; however, this restriction can be circumvented in non-minimally coupled models, leading to physically viable hairy BH solutions. In the framework of Einstein-conformally coupled scalar theory, the Bocharova-Bronnikov-Melnikov-Bekenstein (BBMB) BH was discovered as an exact solution featuring a mass $M$ and a non-trivial scalar field $\phi(r)$, which diverges at the event horizon \cite{Bocharova:1970skc,Bekenstein:1974sf}. To address the divergence of the scalar field at the horizon, the inclusion of a negative cosmological constant was proposed, allowing the singularity to be hidden behind the horizon in asymptotically anti-de Sitter spacetimes \cite{Martinez:2002ru,Martinez:2005di}. However, this regularization results in a spacetime that is no longer asymptotically flat. A significant advancement was achieved by extending the theory to include a Maxwell field, resulting in the Einstein-Maxwell-Conformally coupled Scalar (EMCS) theory. Within this framework, Astorino constructed a charged, spherically symmetric BH solution with a regular scalar field at the horizon \cite{Astorino:2013sfa}, representing a crucial step toward physically realistic violations of the no-hair theorem. In this work, we employ the NJA to construct rotating charged BH solutions with scalar hair, thereby enabling a systematic investigation of thin accretion disk physics and BH shadows in such spacetimes. We examine how the spin parameter $a$, electric charge $Q$, and scalar hair parameter $s$ collectively shape observable signatures—including the energy flux, radiation temperature, and shadow angular size. By confronting our theoretical predictions with EHT observations of Sgr A$^*$, we place constraints on the physically allowed parameter space of these hairy BHs, offering new insights into potential deviations from GR in the strong-field regime.
	
	The structure of this paper is as follows. In Sec.~\ref{section2}, we begin with a review of the static charged BH solution with scalar hair in the EMCS theory, and subsequently apply the NJA to construct its rotating generalization. We also analyze its horizon structure and ergosurface. In Sec.~\ref{section3}, we investigate the energy flux and temperature distribution of thin accretion disks around the rotating charged BH solution with scalar hair. In Sec.~\ref{section4}, we analyze the geodesic equations of photons to study the shadow cast by this rotating charged BH with scalar hair. In Sec.~\ref{section5}, we use BH shadow observables to constrain the parameters $Q$ and $s$, employing EHT observations of Sgr~A$^*$. Finally, we present our conclusions in Sec.~\ref{section6}.
	
	\section{rotating charged black hole with scalar hair}\label{section2}
	\subsection{Construction of rotating charged black hole with scalar hair}
	A well-known violation of the classical no-hair theorem arises in EMCS theory, which admits an exact static, charged BH solution with nontrivial scalar hair \cite{Martinez:2005di}. The theory is defined by the action
	\begin{equation}
		S = \frac{1}{16\pi G}\int d^4x \sqrt{-g} \left[R - F_{\mu\nu}F^{\mu\nu} - 8\pi G\left(\nabla_{\mu}\psi\nabla^{\mu}\psi+\frac{R}{6}\psi^{2}\right)  \right] ,
		\label{action}
	\end{equation}
	where $R$ is the Ricci scalar, $F^{\mu\nu}$ is the electromagnetic field strength and $\psi$ is a scalar field. This coupling allows for regular scalar hair outside the event horizon, in contrast to minimally coupled models. The corresponding static and spherically symmetric BH solution is obtained by solving the field equations derived from action (\ref{action}). The line element takes the form \cite{Astorino:2013sfa}
	\begin{equation}
		d s^{2}=-f(r) d t^{2}+ \frac{1}{f(r)} d r^{2}+r^{2}\left(d \theta^{2}+\sin ^{2} \theta d \phi^{2}\right),\label{xianyuan}
	\end{equation}
	and
	\begin{equation}
		f(r)=1-\frac{2 M}{r}+\frac{Q^{2}+s}{r^{2}},
		\label{dugui}
	\end{equation}
	where $M$, $Q$, and $s$ are the BH mass, electric charge, and scalar hair parameter, respectively.
	
	Using the NJA \cite{Newman:1965tw}, we construct the metric for a rotating charged BH with scalar hair in Boyer-Lindquist coordinates. The resulting line element reads:
	\begin{equation}
		\begin{split}
			ds^2 &= -g_{tt}dt^{2} + g_{rr}dr^{2} + g_{\theta\theta}d\theta^{2} + g_{\phi\phi}d\phi^{2} + 2g_{t\phi}dtd\phi \\
			&= -\left(1-\frac{2\tilde{M} r}{\rho^{2}}\right)d t^{2} + \frac{\rho^{2}}{\Delta}d r^{2} + \rho^{2}d \theta^{2}+ \frac{\Sigma\sin^{2}\theta}{\rho^{2}}d\phi^{2}\\
			&\quad - \frac{4a\tilde{M} r \sin^{2}\theta}{\rho^{2}}dtd\phi,\label{xianyuan2}
		\end{split}
	\end{equation}
	where
	\begin{equation}
		\begin{split}
			&\tilde{M}=\frac{2Mr-Q^{2}-s}{2r},\\
			&\rho^{2}=r^{2}+a^{2}\cos^{2}\theta,\\
			&\Delta=a^2+r^{2} \left(1-\frac{2 M}{r}+\frac{Q^{2}+s}{r^{2}}\right),\\
			&\Sigma=\left(a^{2}+r^{2}\right)^2-a^2 \sin ^2\theta \left(a^{2}+r^{2} \left(1-\frac{2 M}{r}+\frac{Q^{2}+s}{r^{2}}\right)\right).\label{dugui2}
		\end{split}
	\end{equation}
	Here, $a$ is the spin parameter of the BH. In the limiting cases, the above metric reduces to well-known solutions: it describes a Kerr BH when $Q \rightarrow 0$ and $s \rightarrow 0$, and further reduces to the Schwarzschild solution when the spin $a$ is also set to zero. 
	\subsection{Properties of the rotating charged black holes with scalar hair}
\begin{figure}[htbp]
		\centering
		\begin{subfigure}{0.45\textwidth}
			\includegraphics[width=3.2in, height=5.5in, keepaspectratio]{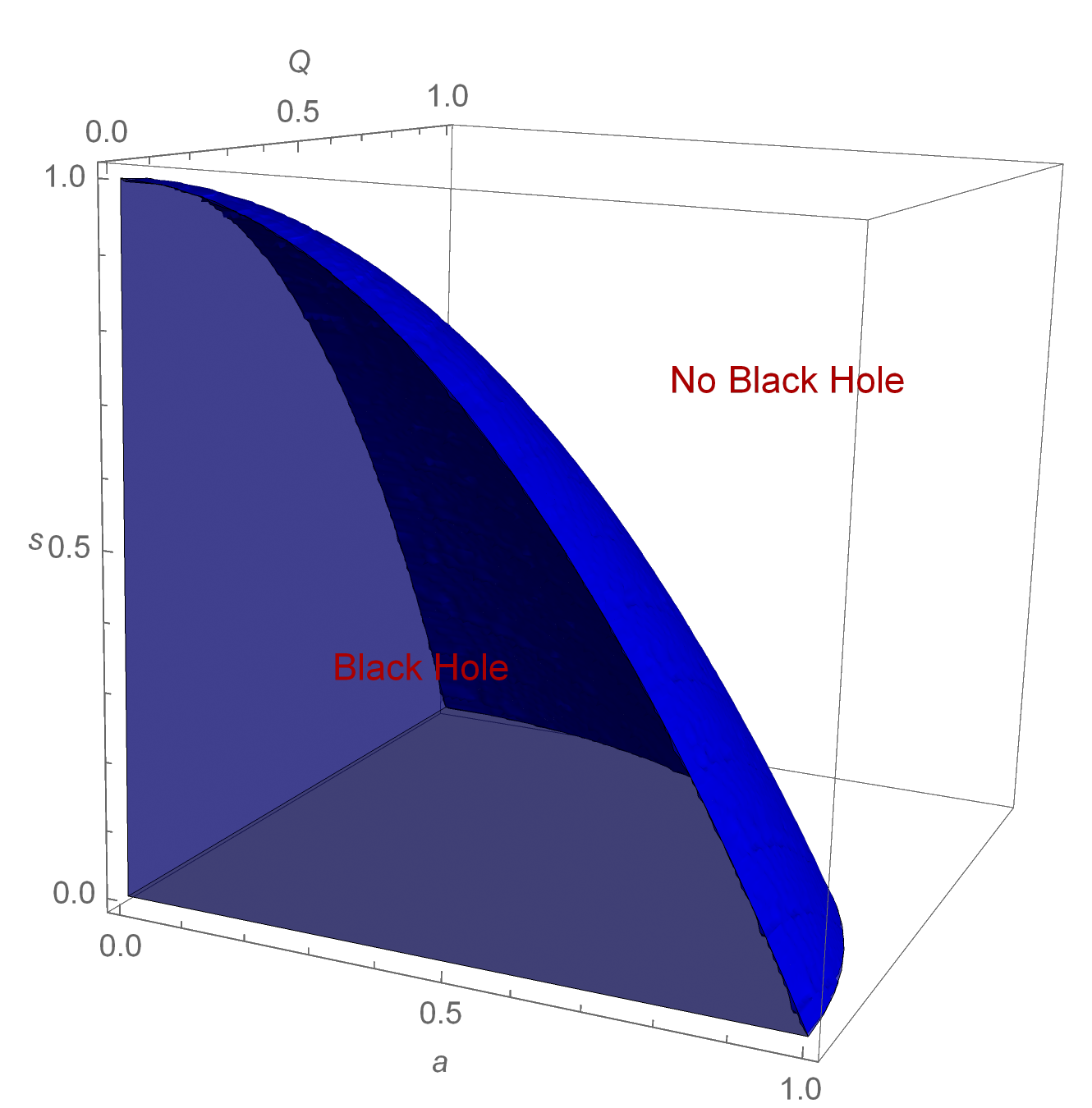}
		\end{subfigure}
		\caption{Regions of BH horizon existence in the $(a, Q, s)$ parameter space.}
		\label{horizon}
	\end{figure}

\begin{figure*}[htbp]
		\centering
		
		\begin{subfigure}{0.3\textwidth}
			\centering
			\includegraphics[width=\linewidth, keepaspectratio]{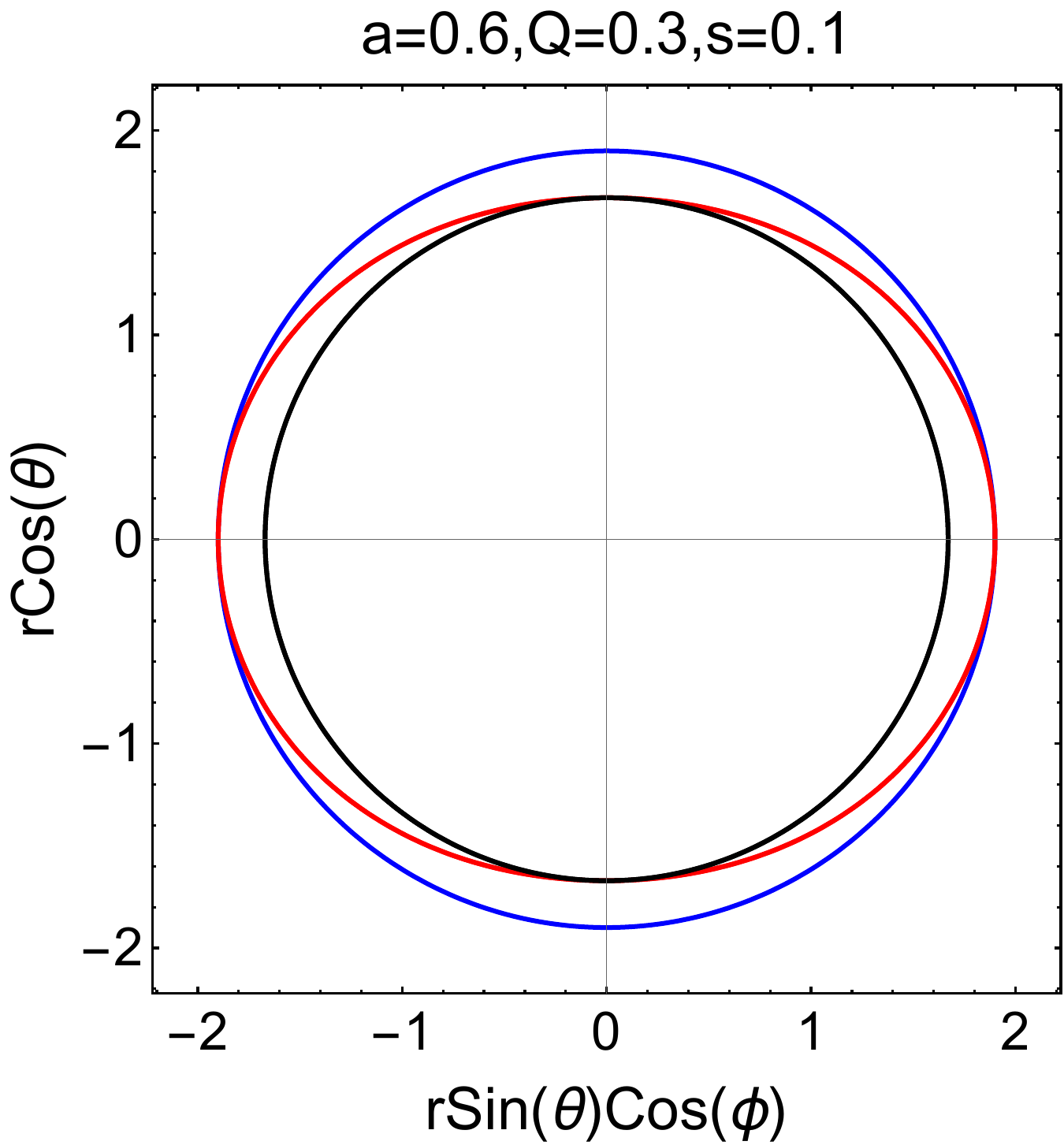}
		\end{subfigure}
		\hfill
		\begin{subfigure}{0.3\textwidth}
			\centering
			\includegraphics[width=\linewidth, keepaspectratio]{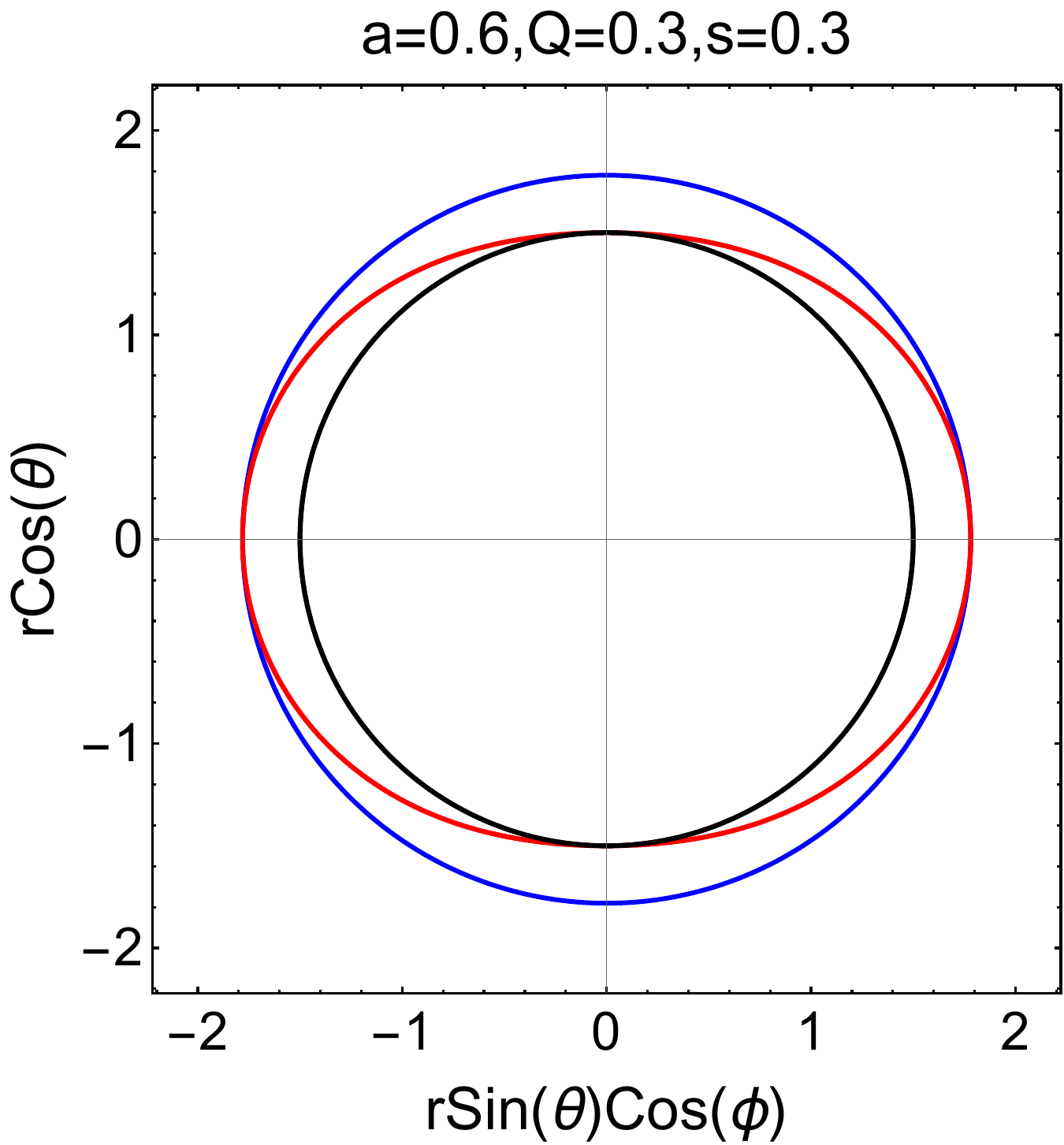}
		\end{subfigure}
		\hfill
		\begin{subfigure}{0.3\textwidth}
			\centering
			\includegraphics[width=\linewidth, keepaspectratio]{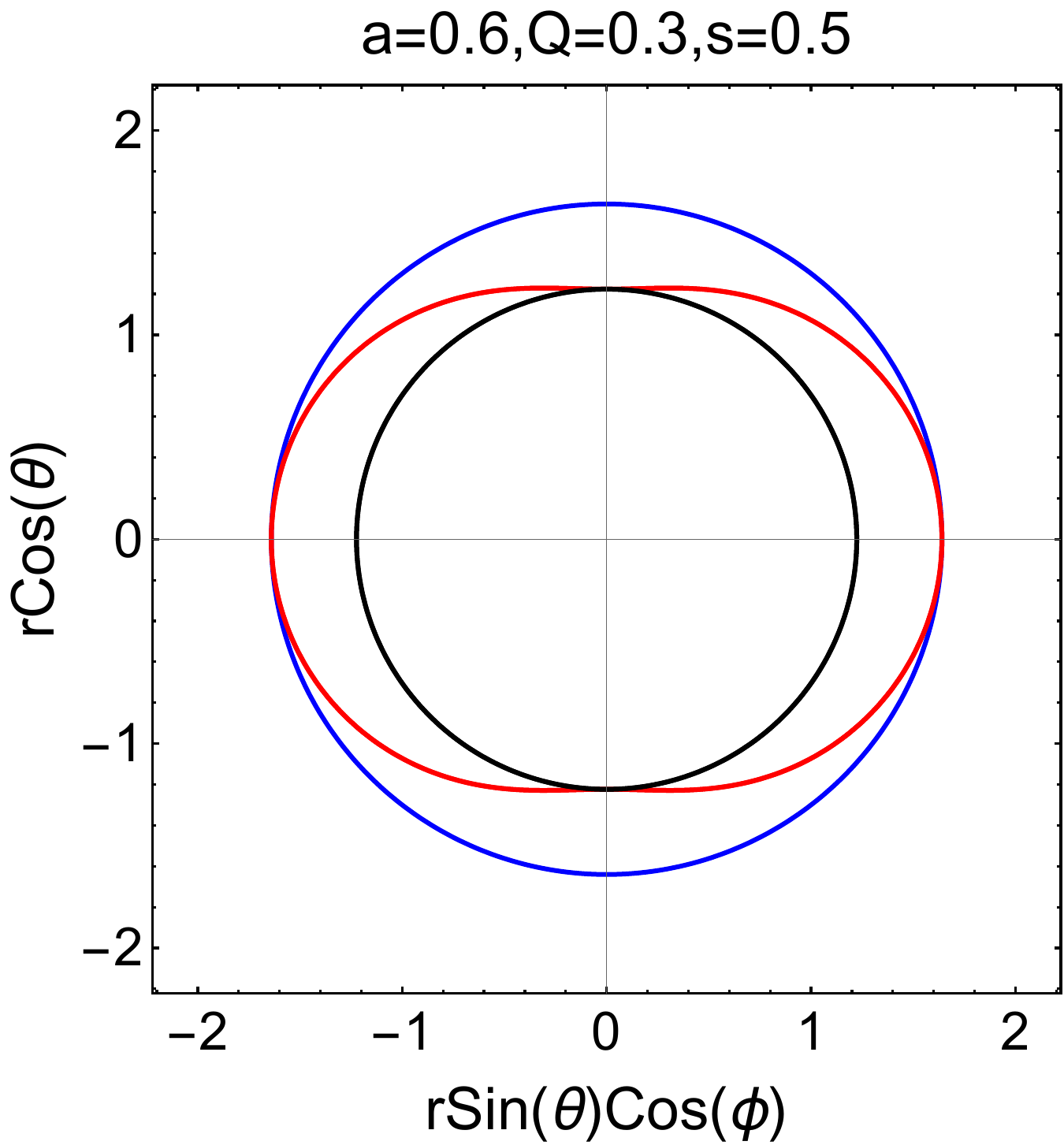}
		\end{subfigure}
		
		\vspace{0.5cm}

		\begin{subfigure}{0.3\textwidth}
			\centering
			\includegraphics[width=\linewidth, keepaspectratio]{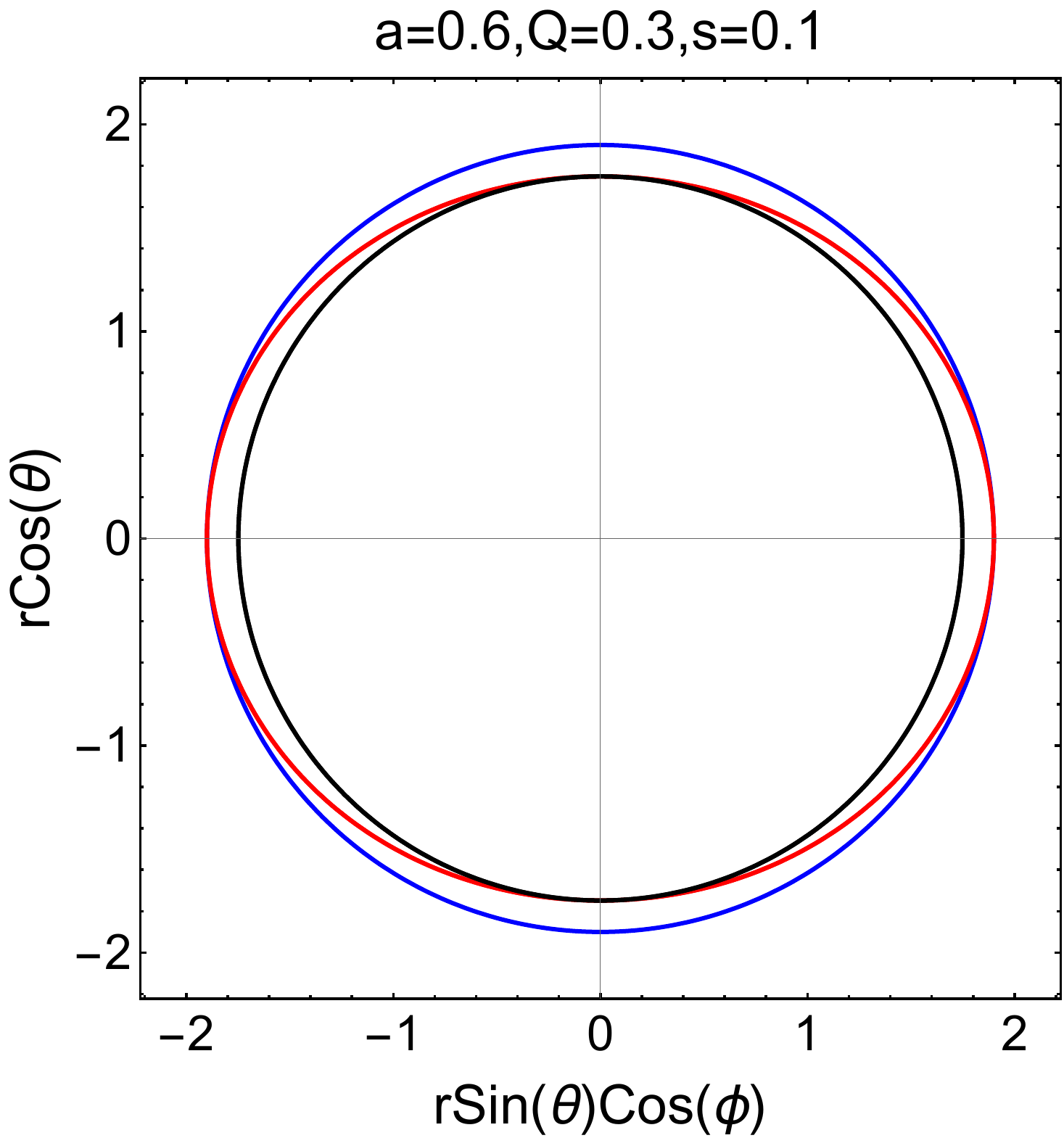}
		\end{subfigure}
		\hfill
		\begin{subfigure}{0.3\textwidth}
			\centering
			\includegraphics[width=\linewidth, keepaspectratio]{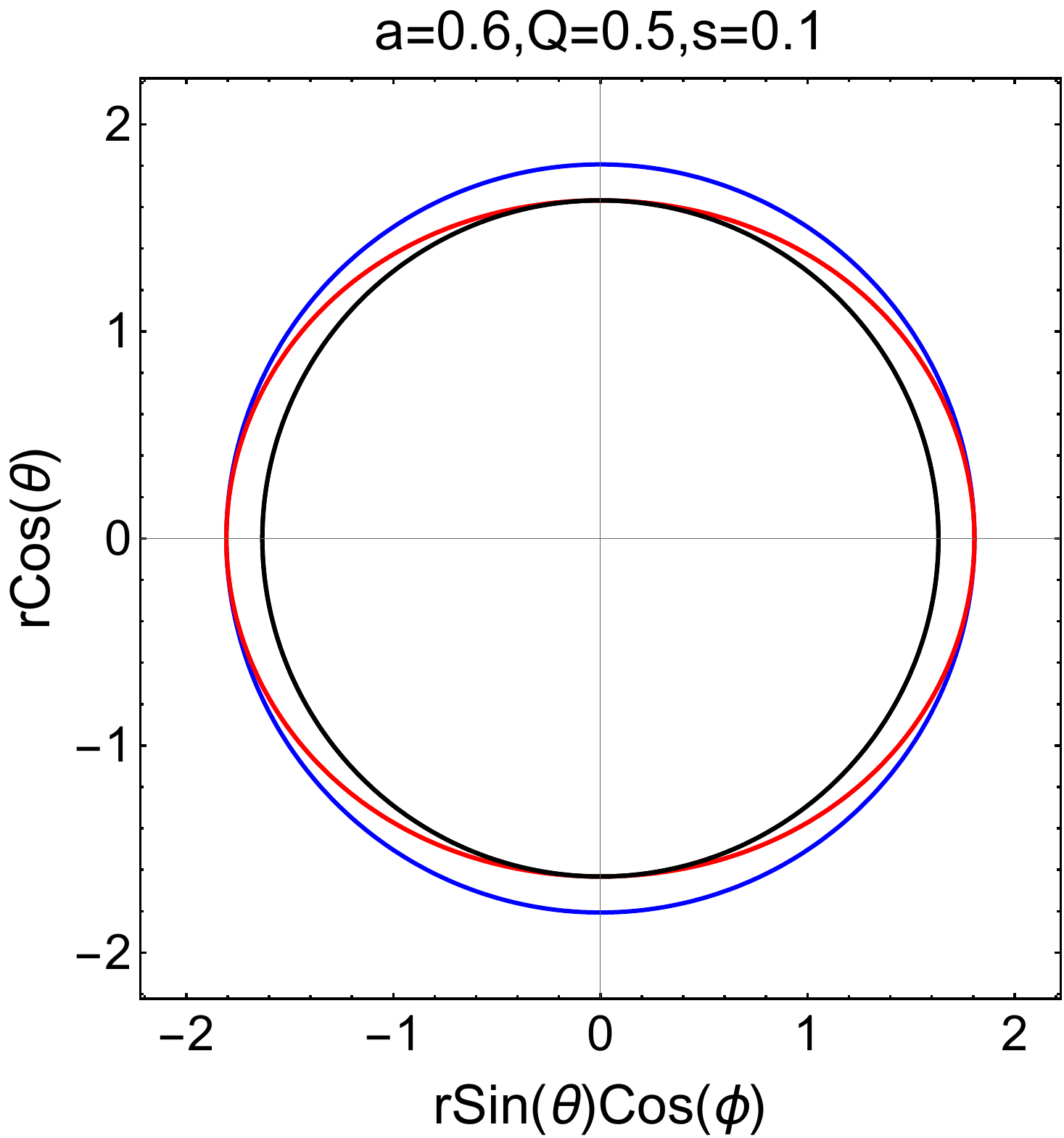}
		\end{subfigure}
		\hfill
		\begin{subfigure}{0.3\textwidth}
			\centering
			\includegraphics[width=\linewidth, keepaspectratio]{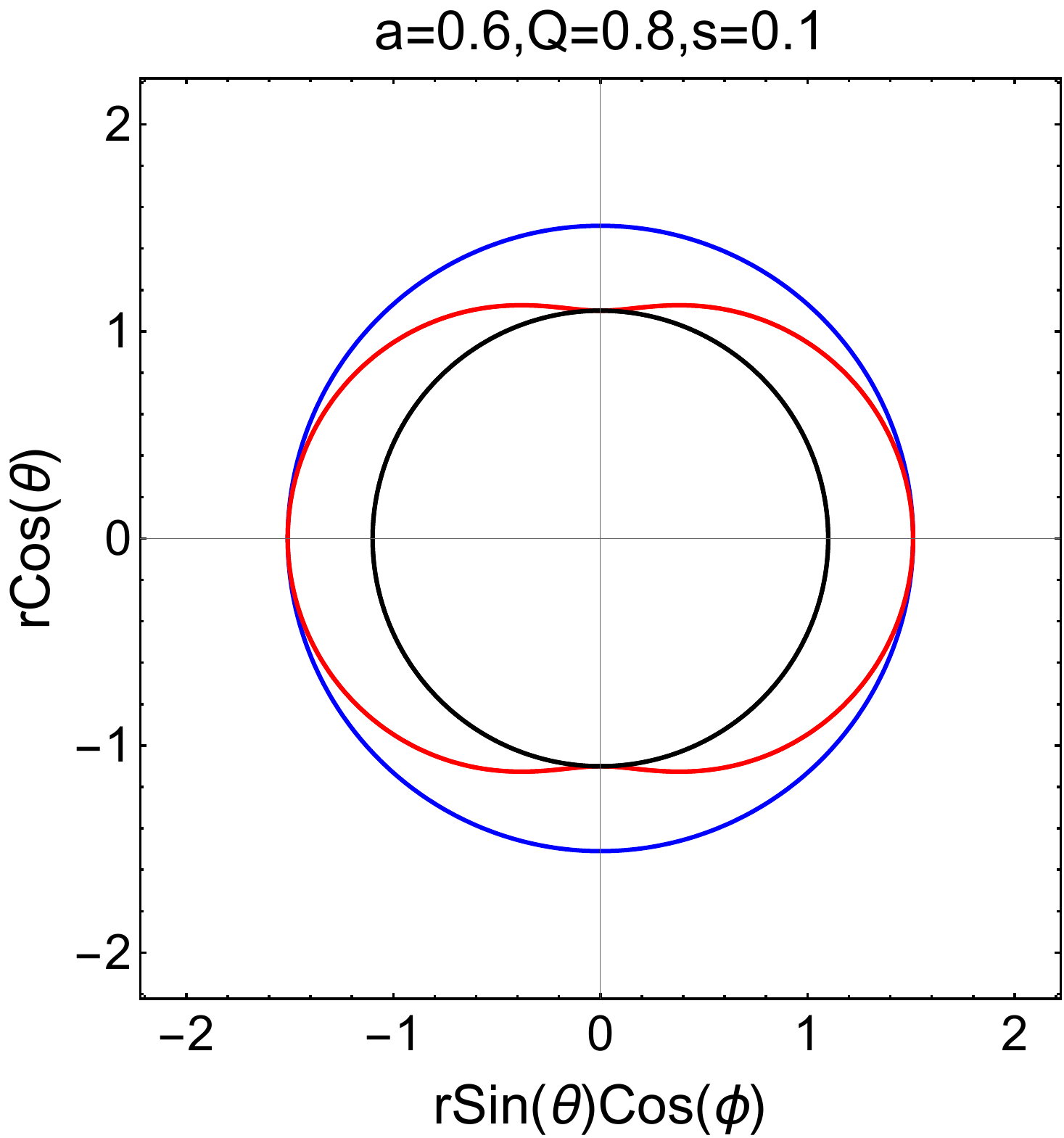}
		\end{subfigure}
		
		\caption{The shapes of the outer horizon (black solid line) and outer ergosphere (red solid line) are compared to the outer event
			horizon in the static case (blue solid line).}
		\label{surface}
	\end{figure*}

	Next, we investigate the physical properties of rotating charged BHs with scalar hair. The horizons of rotating charged BHs with scalar hair are determined by the condition $g^{rr}=0$, which yields the following expression:
	\begin{equation}
		\begin{split}
			\Delta=a^{2}+r^{2} \left(1-\frac{2 M}{r}+\frac{Q^{2}+s}{r^{2}}\right)=0.\label{data}
		\end{split}
	\end{equation}
	The horizon radii are determined by the roots of Eq. (\ref{data}), which exhibit significant sensitivity to the parameters $M$, $a$, $Q$ and $s$. This equation yields two distinct real roots, corresponding to the inner and outer event horizons, respectively. Figure~\ref{horizon} illustrates the parameter space $(a, Q, s)$ for rotating charged BHs with scalar hair, indicating the region in which physically viable BH solutions exist. The blue solid surface represents the extremal limit of black hole configurations, while the light blue region enclosed by it delineates the domain in which regular black hole solutions exist.
	The two-dimensional surfaces satisfying $g_{tt} = 0$ correspond to the inner and outer ergosurfaces, leading to
	\begin{equation}
		\begin{split}
			r^{2} \left(1-\frac{2 M}{r}+\frac{Q^{2}+s}{r^{2}}\right)+a^{2}\cos^{2}\theta=0.\label{ergosurface}
		\end{split}
	\end{equation}
	Figure~\ref{surface} illustrates the variations in the outer horizon and outer ergosurface across different parameter values. A blue circle represents the static scenario where $a=0$. Specifically, when holding $a$ and $Q$ constant, both the radii of the outer ergosphere and the outer horizon diminish as $s$ increases. Similarly, for a given $a$ and $s$, these radii decrease with an increase in $Q$.
	
	\section{The physical properties of thin accretion disks}\label{section3}
	In this section, we begin with a brief overview of the physical properties of thin accretion disks as described by the Novikov-Thorne model \cite{NT}. Subsequently, we examine how the parameters $Q$ and $s$ influence the energy flux and radiation temperature in detail. The following values are adopted for the physical constants and the characteristics of the thin accretion disk in our analysis: $c=2.997\times10^{10}\text{cms}^{-1}$, $\dot{M}_{0}=2\times10^{-6}M_{\odot}\text{yr}^{-1}$, $1\text{yr}=3.156\times10^{7}\text{s}$, $\sigma_{\text{SB}}=5.67\times10^{-5}\text{ergs}^{-1}\text{cm}^{-2}\text{K}^{-4}$, $h=6.625\times10^{-27}\text{ergs}$, $k_{\text{B}}=1.38\times10^{-16}\text{erg}\text{K}^{-1}$, $M_{\odot}=1.989\times10^{33}\text{g}$, and the mass of BH $M=2\times10^{6}M_{\odot}$.
	In the equatorial plane ($\theta=\pi/2$), we analyze the geodesics of radial timelike particles. The corresponding Lagrangian is expressed as:
	\begin{equation}
		\mathcal{L}=-g_{tt}\dot{t}^{2}+2g_{t\phi}\dot{t}\dot{\phi}+g_{rr}\dot{r}^{2}+g_{\phi\phi}\dot{\phi}^{2},\label{la}
	\end{equation}
	here, a dot indicates differentiation with respect to the affine parameter $\tau$. Based on the Euler-Lagrange equation $(\frac{d}{d\tau}(\frac{\partial\mathcal{L}}{\partial\dot{x}})-\frac{\partial\mathcal{L}}{\partial x}=0)$, for variables $x = t, \phi$, it follows that
	\begin{equation}
		\begin{split}
			-g_{tt}\dot{t} + g_{t\phi}\dot{\phi} &= -E, \\
			g_{t\phi}\dot{t} + g_{\phi\phi}\dot{\phi} &= L.\label{EL}
		\end{split}
	\end{equation}
	The constants $E$ and $L$ represent the specific energy and angular momentum of particles moving in circular orbits around compact objects. By solving the aforementioned equations, we obtain the four-velocity components of the orbiting particles as follows:
	\begin{equation}
		\begin{split}
			\dot{t} &=\frac{Eg_{\phi\phi}+Lg_{t\phi}}{g_{t\phi}^{2}-g_{tt}g_{\phi\phi}}, \\
			\dot{\phi} &=-\frac{Eg_{t\phi}+Lg_{tt}}{g_{t\phi}^{2}-g_{tt}g_{\phi\phi}}.\label{tfai}
		\end{split}
	\end{equation}
	Based on the normalization condition $g_{\mu\nu}\dot{x}^{\mu}\dot{x}^{\nu}=-1$, it follows that 
	\begin{equation}
		g_{rr}\dot{r}^{2}=V_{eff}.\label{Veff}
	\end{equation}
	The effective potential takes the form
	\begin{equation}
		V_{eff}=-1+\frac{E^{2}g_{\phi\phi}+2ELg_{t\phi}+L^{2}g_{tt}}{g_{t\phi}^{2}-g_{tt}g_{\phi\phi}}.\label{Veff2}
	\end{equation}
	In the equatorial plane ($\theta=\frac{\pi}{2}$), circular orbits are identified by the requirements $V_{eff}(r)=0$ and $V_{eff,r}(r)=0$. These constraints allow us to express the specific energy $E$ and specific angular momentum $L$ in terms of the particles' angular velocity $\Omega$.
	\begin{equation}
		\Omega=-\frac{d\phi}{dt}=\frac{-g_{t\phi,r}+\sqrt{(g_{t\phi,r})^{2}-g_{tt,r}g_{\phi\phi,r}}}{g_{\phi\phi,r}},\label{11}
	\end{equation}
	\begin{equation}
		E=-\frac{g_{tt}+g_{t\phi}\Omega}{\sqrt{-g_{tt}-2g_{t\phi}\Omega-g_{\phi\phi}\Omega^{2}}},\label{12}
	\end{equation}	
	\begin{equation}
		L=\frac{g_{t\phi}+g_{\phi\phi}\Omega}{\sqrt{-g_{tt}-2g_{t\phi}\Omega-g_{\phi\phi}\Omega^{2}}}.\label{13}
	\end{equation}
	To establish the inner edge of the disk, we need to compute the innermost stable circular orbit (ISCO) of the BH potential based on the condition $V_{eff,rr}(r)=0$. This condition leads to the following relation:
	\begin{equation}
		E^{2}g_{\phi\phi,rr}+2ELg_{t\phi,rr}+L^{2}g_{tt,rr}-(g_{t\phi}^{2}-g_{tt}g_{\phi\phi})_{,rr}=0.\label{14}
	\end{equation}
	The radius $r_{isco}$ marks the inner edge of the thin accretion disk, as equatorial circular orbits for $r<r_{isco}$ are unstable.
	
\begin{figure*}[htbp]
		\centering
		\begin{subfigure}{0.45\textwidth}
			\includegraphics[width=3in, height=5.5in, keepaspectratio]{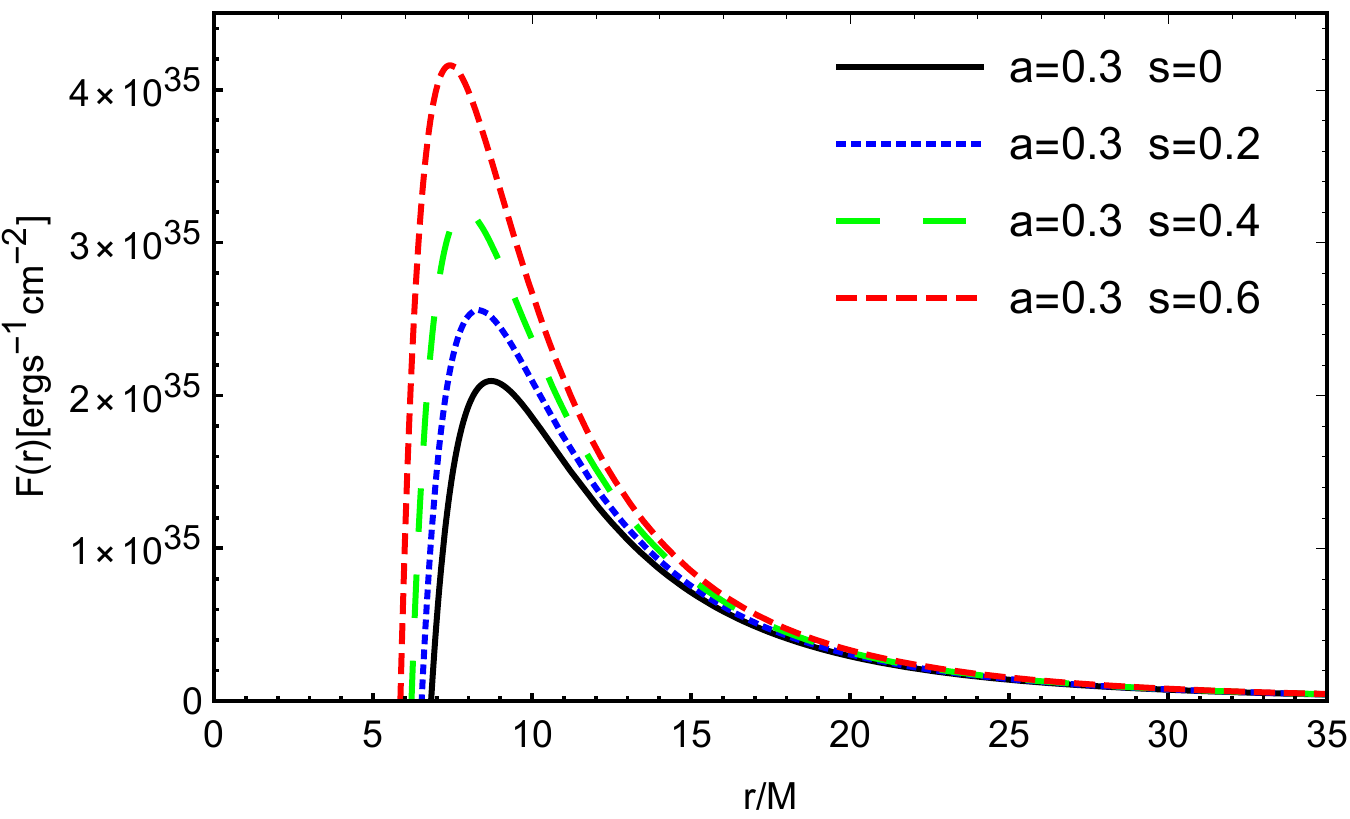}
		\end{subfigure}
		\hfill
		\begin{subfigure}{0.45\textwidth}
			\includegraphics[width=3in, height=5.5in,keepaspectratio]{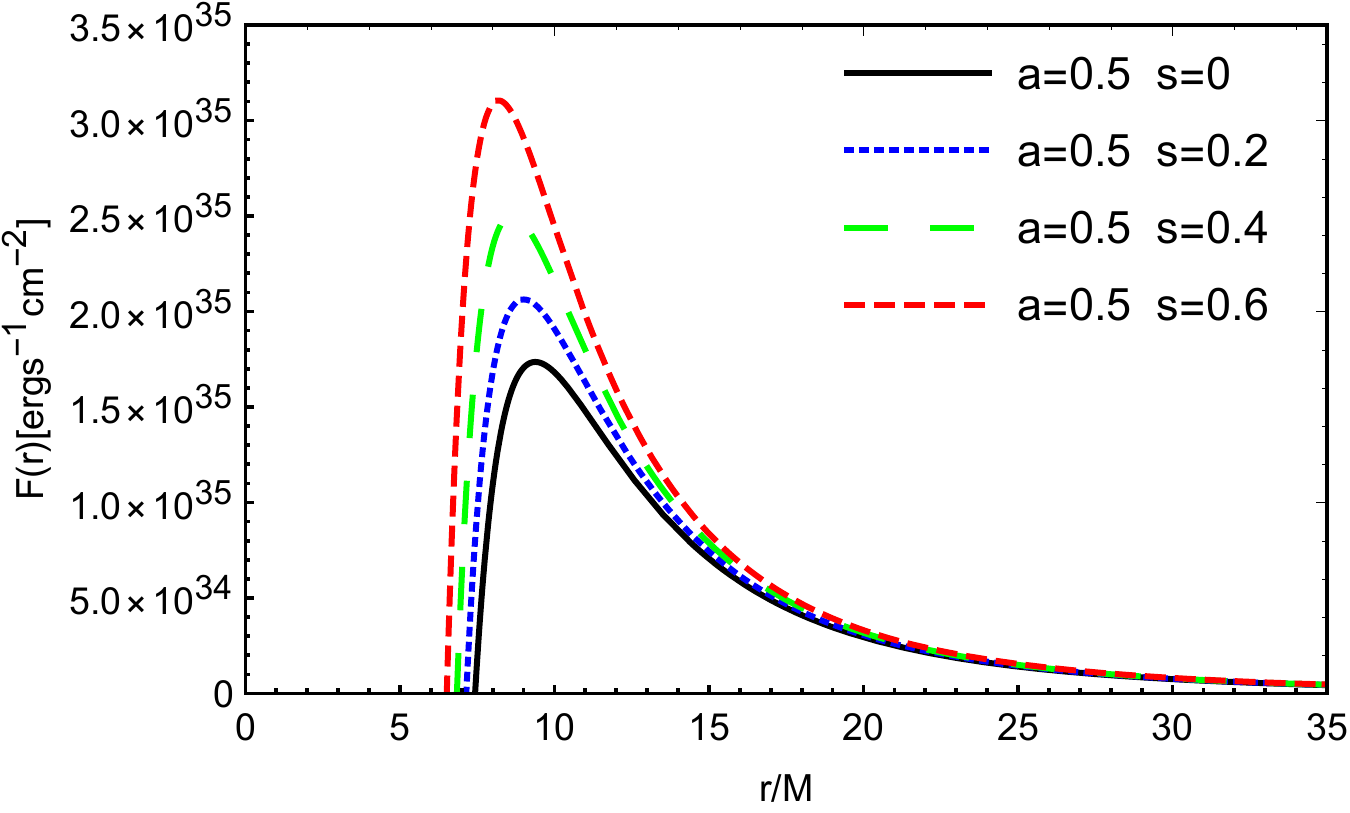}
		\end{subfigure}
		\begin{subfigure}{0.45\textwidth}
			\includegraphics[width=3in, height=5.5in, keepaspectratio]{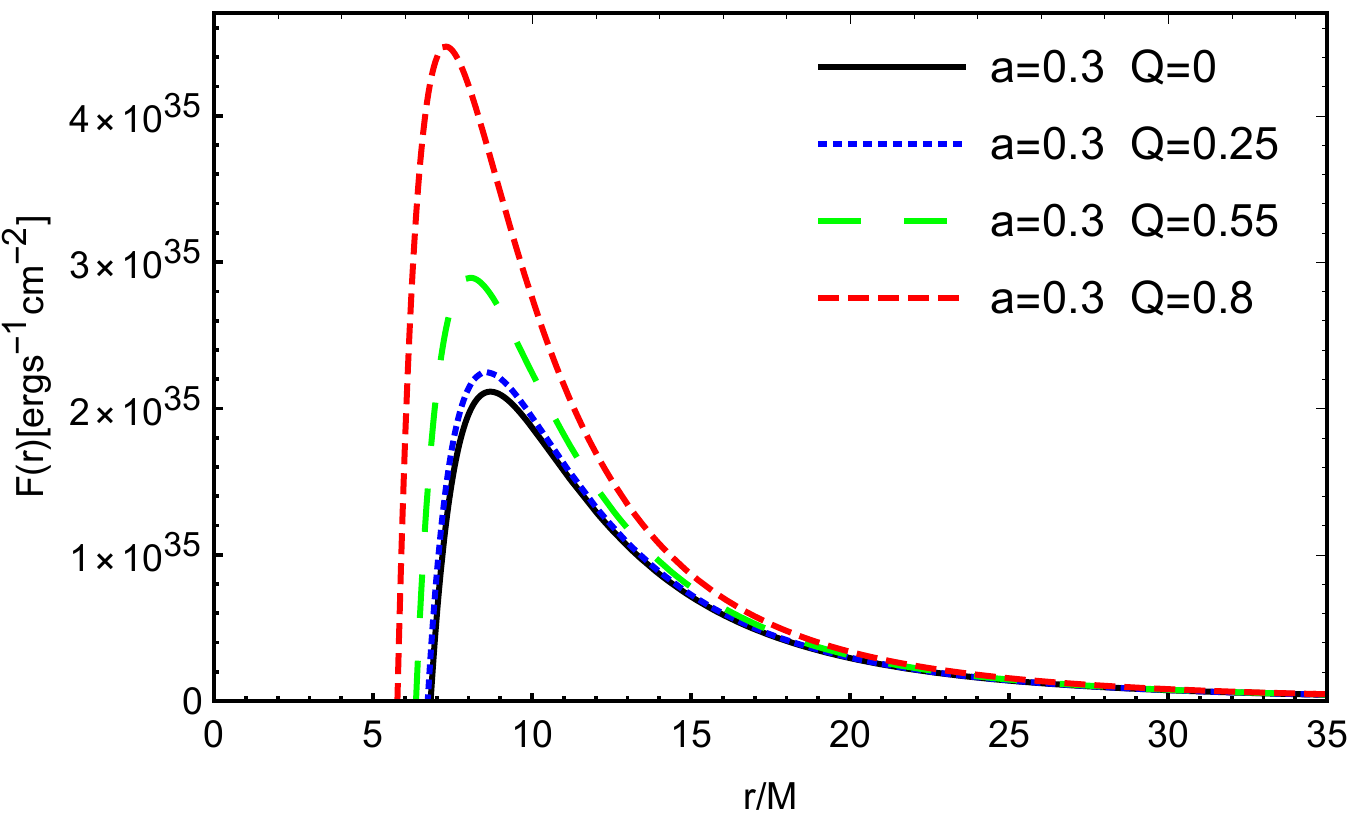}
		\end{subfigure}
		\hfill
		\begin{subfigure}{0.45\textwidth}
			\includegraphics[width=3in, height=5.5in, keepaspectratio]{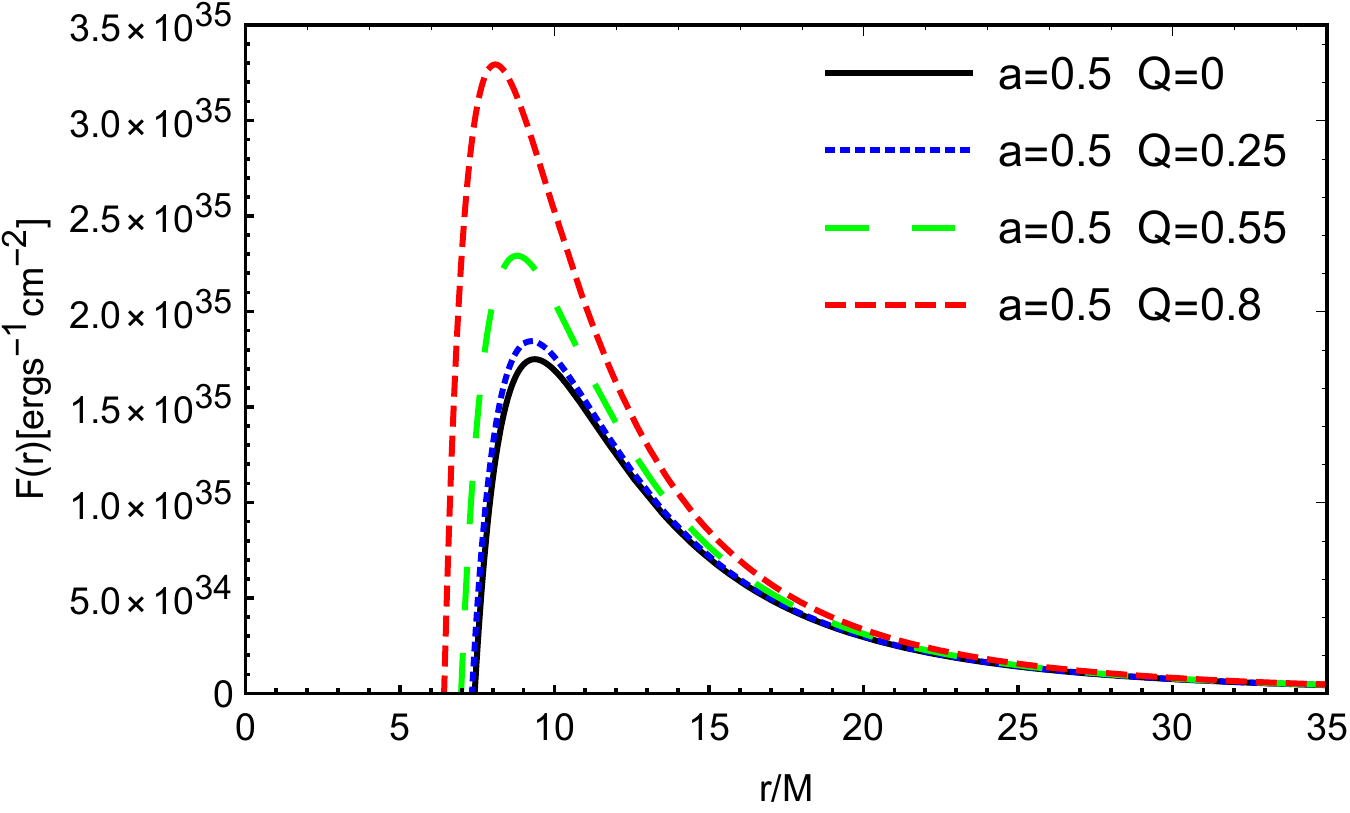}
		\end{subfigure}
		\caption{The energy flux $F(r)$ from an accretion disk around a rotating charged black hole with scalar hair is plotted for different values of $Q$ and $s$. The top row shows results for fixed $Q = 0.3$, while the bottom row corresponds to fixed $s = 0.1$.}
		\label{Fr}
	\end{figure*}

	Next, we investigate accretion processes in thin disks surrounding rotating charged BHs with scalar hair. The radiation flux emitted from the disk surface is governed by three fundamental equations describing the conservation of rest mass, energy, and angular momentum of the accreting particles~\cite{Page:1974he,NT}.
	\begin{equation}
		F(r)=-\frac{\dot{M}_{0}\Omega_{,r}}{4\pi\sqrt{-g}(E-\Omega L)^{2}}\int^{r}_{r_{isco}}(E-\Omega L)L_{,r}dr.\label{15}
	\end{equation}
	The equation is commonly used in the literature for cylindrical coordinates. However, for applications in spherical coordinates, it must be reformulated as detailed in \cite{Collodel:2021gxu}
	\begin{equation}
		F(r)=-\frac{\dot{M}_{0}\Omega_{,r}}{4\pi\sqrt{-g/g_{\theta\theta}}(E-\Omega L)^{2}}\int^{r}_{r_{isco}}(E-\Omega L)L_{,r}dr,\label{16}
	\end{equation}
	where $\dot{M}_{0}$ stands for the mass accretion rate, $g$ refers to the metric determinant. Figure~\ref{Fr} shows the radial dependence of the energy flux $F(r)$ emitted from the accretion disk around a rotating charged BH with scalar hair, for different values of the parameters $Q$ and $s$. The flux initially increases with radius, reaches a peak, and then gradually decreases. For a fixed spin parameter $a$, both the magnitude and peak position of $F(r)$ shift upward as either $Q$ or $s$ increases. In contrast, at fixed $Q$ or $s$, the maximum energy flux diminishes with increasing $a$.
	
\begin{figure*}[htbp]
	\centering
	\begin{subfigure}{0.45\textwidth}
		\includegraphics[width=3in, height=5.5in, keepaspectratio]{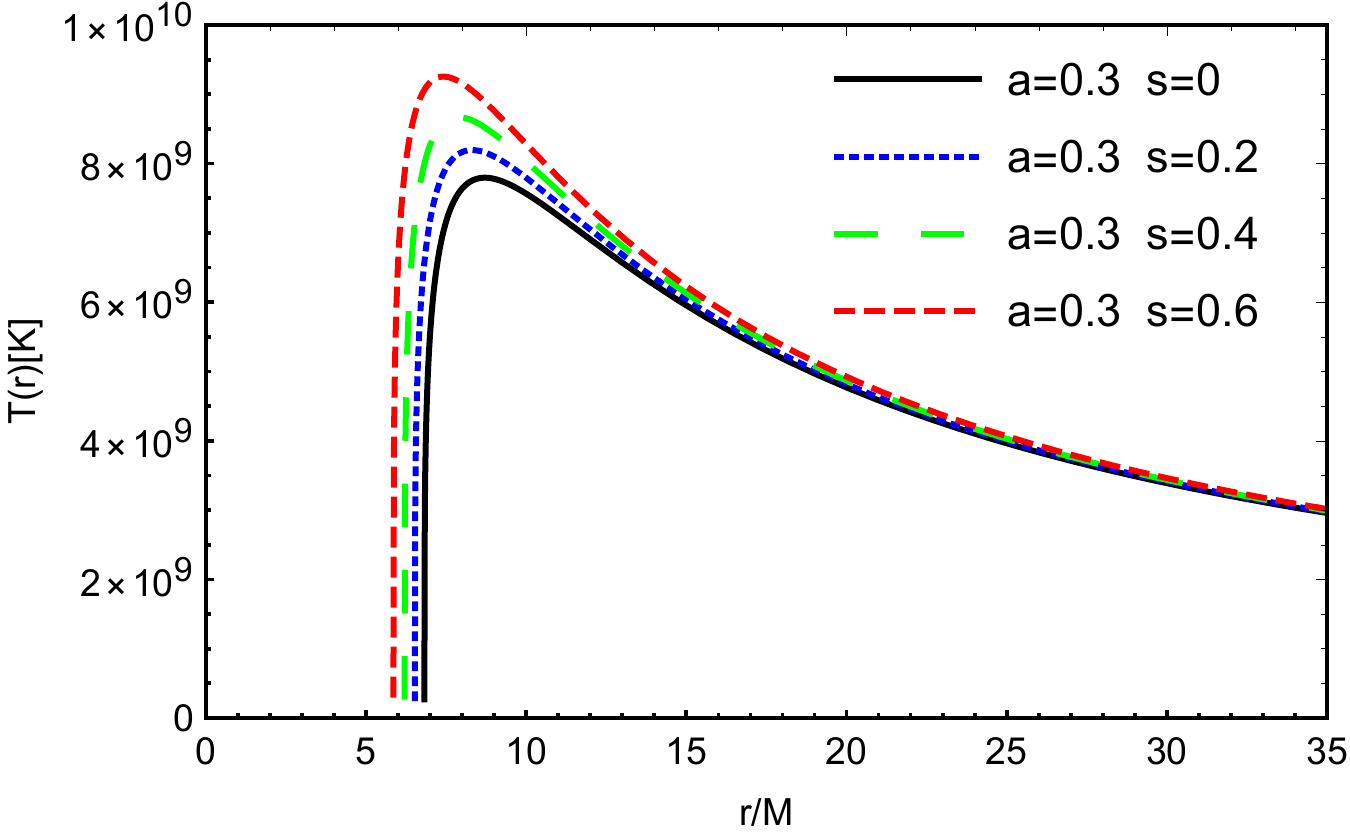}
	\end{subfigure}
	\hfill
	\begin{subfigure}{0.45\textwidth}
		\includegraphics[width=3in, height=5.5in,keepaspectratio]{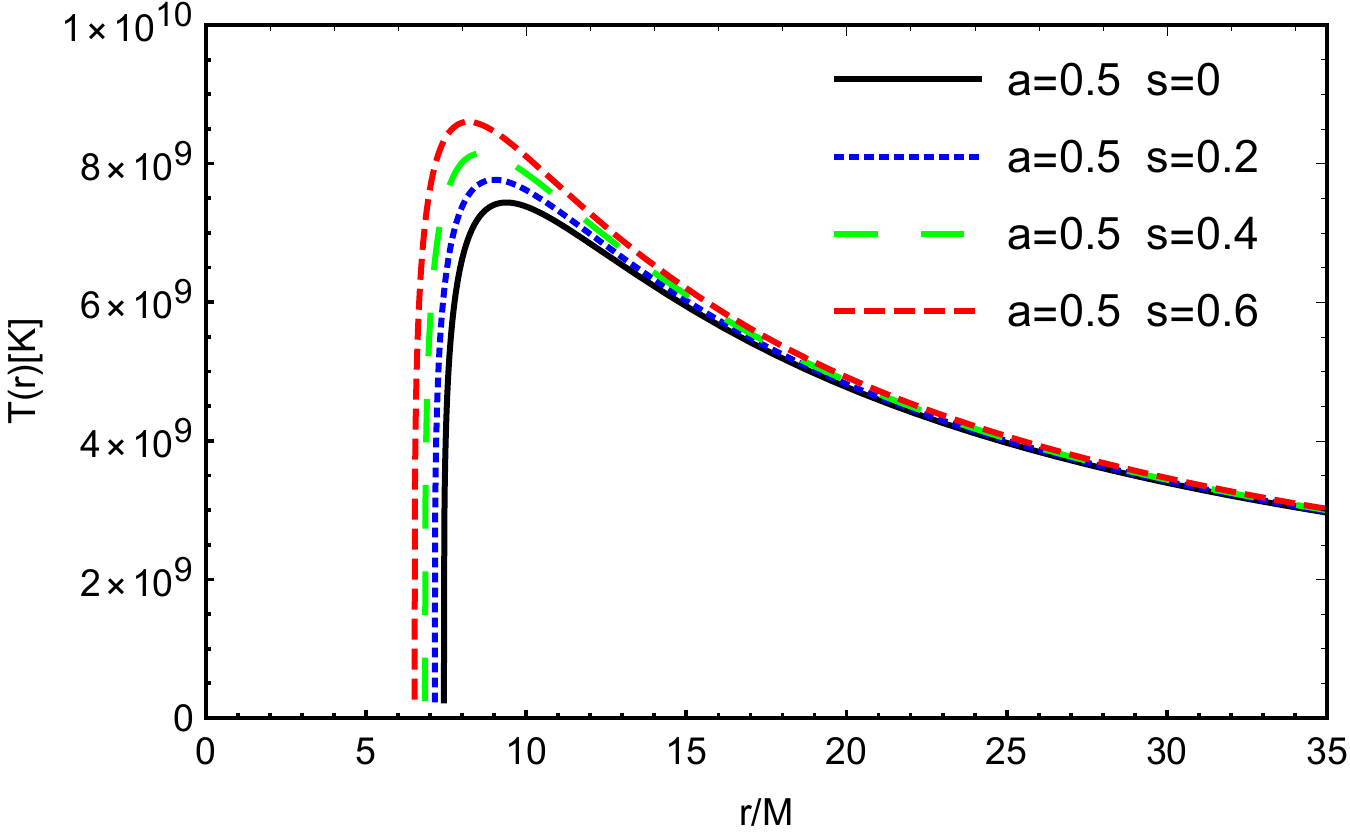}
	\end{subfigure}
	\begin{subfigure}{0.45\textwidth}
		\includegraphics[width=3in, height=5.5in, keepaspectratio]{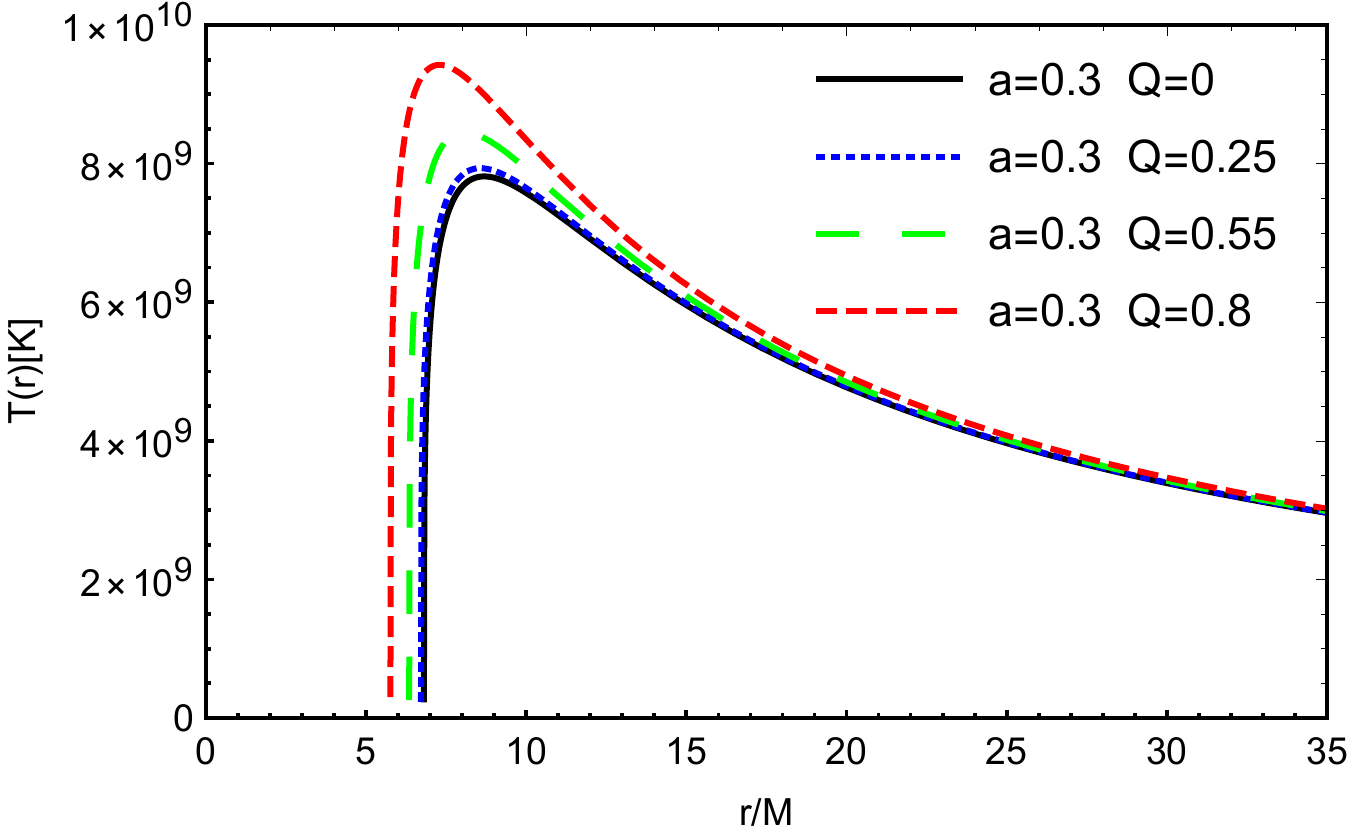}
	\end{subfigure}
	\hfill
	\begin{subfigure}{0.45\textwidth}
		\includegraphics[width=3in, height=5.5in, keepaspectratio]{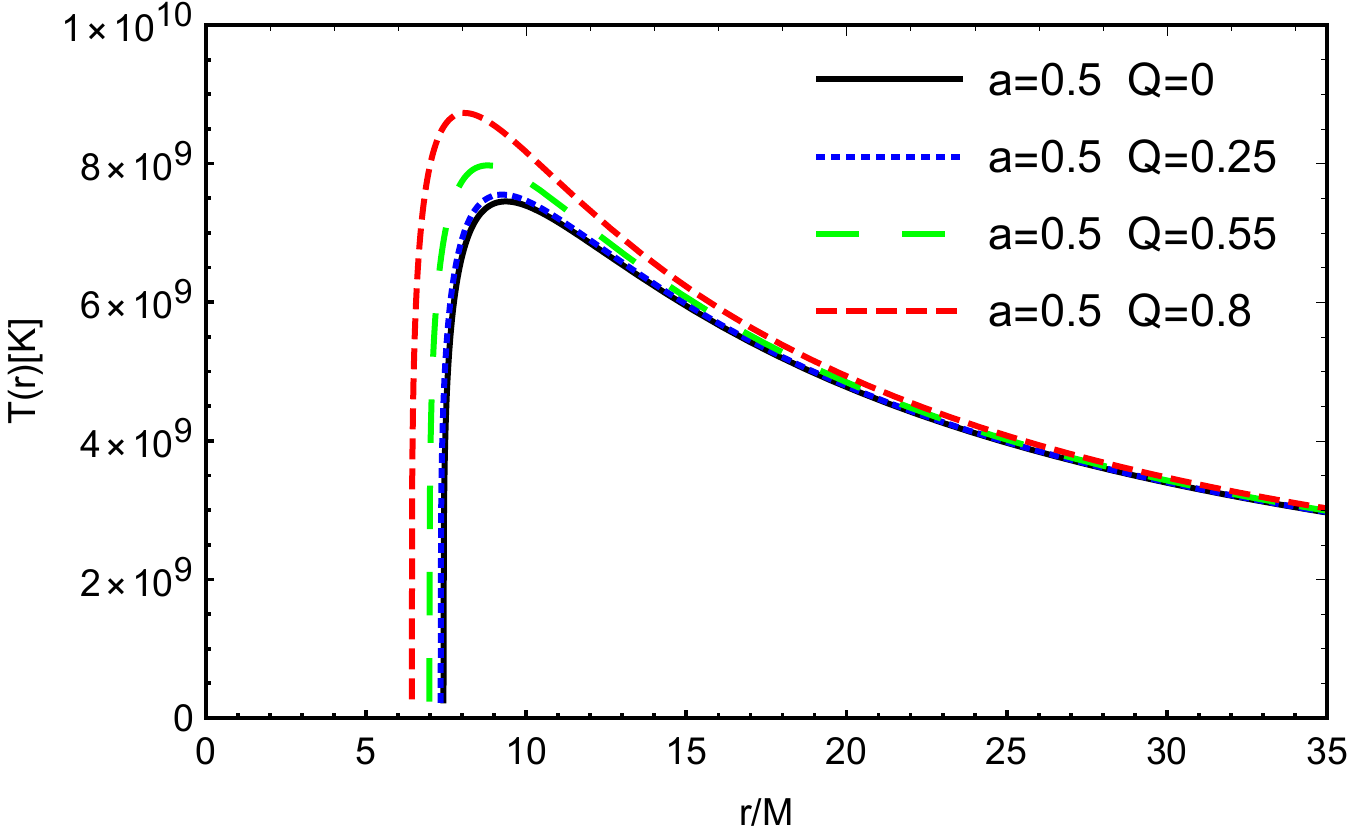}
	\end{subfigure}
	\caption{The temperature $T(r)$ from an accretion disk around a rotating charged black hole with scalar hair is shown for various values of $Q$ and $s$. The top row presents results for a fixed value of $Q=0.3$, whereas the bottom row corresponds to a fixed value of $s=0.1$.}
	\label{Tr}
    \end{figure*}

	Within the Novikov-Thorne model framework, the accreted material reaches a state of thermodynamic equilibrium, which implies that the radiation emitted by the disk closely resembles that of a perfect black body. The relationship between the disk's radiation temperature $T(r)$ and its energy flux $F(r)$ is given by the Stefan-Boltzmann law: $F(r)=\sigma_{SB}T^{4}(r)$, where $\sigma_{SB}$ represents the Stefan-Boltzmann constant. Figure~\ref{Tr} depicts the radiation temperature on the disk surface for different values of the parameters. Similar to the behavior observed for the energy flux, the radiation temperature $T(r)$ follows a characteristic radial profile: it initially rises with radius, reaches a peak at an intermediate distance, and subsequently declines. For a fixed spin parameter $a$, both the magnitude and peak position of $T(r)$ increase as either $Q$ or $s$ increases. Conversely, when $Q$ or $s$ is held constant, the peak temperature decreases with increasing $a$.
	
	\section{Null geodesic equations and black hole shadow}\label{section4}
	
    To determine the shadow cast by a rotating charged BH with scalar hair, we study the null geodesics in this spacetime using the Hamilton-Jacobi formalism, following Carter \cite{Carter:1968rr}
	\begin{equation}
		\frac{\partial S}{\partial \tau }=-\frac{1}{2} g^{\mu \nu} \frac{\partial S}{\partial x^{\mu}} \frac{\partial S}{\partial x^{\nu}}.\label{19}
	\end{equation}
	Here, $\tau$ represents an affine parameter along the geodesics, and $S$ is the Jacobi action. The Jacobi action $S$ can be expressed in the following separable form:
	\begin{equation}
		S=\frac{1}{2} m^{2} \tau -E t+L \phi+S_{r}(r)+S_{\theta}(\theta).\label{20}
	\end{equation}
	Here, $m$ represents the mass of the particle moving within the BH spacetime, and for a photon, $m=0$. The two functions $S_{r}$ and $S_{\theta}$ depend solely on $r$ and $\theta$, respectively. Combining Eq. (\ref{19}) with Eq. (\ref{20}), we can formulate the geodesic equations as follows:
	\begin{equation}
		\rho^{2}\frac{dt}{d\tau}=a\left(L-aE\sin^{2}\theta\right)+\frac{r^{2}+a^{2}}{\Delta(r)}\left[\left(r^{2}+a^{2}\right)E-aL\right],\label{21}
	\end{equation}
	\begin{equation}
		\rho^{2}\frac{dr}{d\tau}=\sqrt{R(r)},\label{22}
	\end{equation}
	\begin{equation}
		\rho^{2}\frac{d\theta}{d\tau }=\sqrt{\Theta(\theta)},\label{23}
	\end{equation}
	\begin{equation}
		\rho^{2}\frac{d\phi}{d\tau}=\frac{L}{\sin^{2}\theta}-aE+\frac{a}{\Delta(r)}\left[\left(r^{2}+a^{2}\right)E-aL\right],\label{24}
	\end{equation}
	where
	\begin{equation}
		R(r)=\left(\left(r^{2}+a^{2}\right)E-aL\right)^{2}-\Delta(r) \left( C +(L - aE)^2 \right),\label{25}
	\end{equation}
	\begin{equation}
		\Theta(\theta)=C+\left(a^{2}E^{2}-L^{2}\csc^{2}\theta\right)\cos^{2}\theta,\label{26}
	\end{equation}
	with $C$ being the Carter constant \cite{Carter:1968rr}. 
	
	The behavior of a photon's motion is defined by the two impact parameters
	\begin{equation}
		\xi=\frac{L}{E}, \quad \eta=\frac{C}{E^{2}}.\label{27}
	\end{equation}
	To ascertain the shape of the BH shadow, we need to calculate the critical circular orbit for photons, based on the unstable condition
	\begin{equation}
		R(r)|_{r=r_{ps}}=\frac{dR(r)}{dr}|_{r=r_{ps}}=0, \quad \frac{d^{2}R(r)}{dr^{2}}|_{r=r_{ps}}>0,\label{28}
	\end{equation}
	here, $r_{ps}$ denotes the radius of the unstable circular photon orbits. Solving Eq. (\ref{28}) provides the critical values of the impact parameters ($\xi, \eta$) for unstable orbits, which are given by: 
	\begin{equation}
		\xi=\frac{a^{2}+r_{ps}^{2}-\frac{4\Delta(r_{ps})r_{ps}}{\Delta^{\prime}(r_{ps})}}{a},\label{29}
	\end{equation}
	\begin{equation}
		\eta=\frac{-16\Delta(r_{ps})^{2}r_{ps}^{2}-r_{ps}^{4}\Delta^{\prime}(r_{ps})^{2}+8\Delta(r_{ps})r_{ps}\left(2a^{2}r_{ps}+r_{ps}^{2}\Delta^{\prime}(r_{ps})\right)}{a^{2}\Delta^{\prime}(r_{ps})^{2}}.\label{30}
	\end{equation}
	To describe the shadow, we utilize the celestial coordinates $(X,Y)$ as follows:
	\begin{equation}
		X=\lim _{r_{0}\rightarrow\infty}\left(-r_{0}^{2}\sin\theta_{0}\frac{d\phi}{dr}|_{\theta\rightarrow\theta_{0}}\right)=-\xi\csc\theta_{0},\label{31}
	\end{equation}
	\begin{equation}
		Y=\lim _{r_{0}\rightarrow\infty}\left(r_{0}^{2}\frac{d \theta}{d r}|_{\theta\rightarrow\theta_{0}}\right) =\pm\sqrt{\eta+a^{2}\cos^{2}\theta_{0}-\xi^{2}\cot^{2} \theta_{0}},\label{32}
	\end{equation}
	where $r_{0}$ is the distance between the observer and the BH, and $\theta_{0}$ is the inclination angle between the observer's line of sight and the rotational axis of the BH. In the particular case where the observer is situated on the BH's equatorial plane with $\theta_{0}=\pi/2$, we obtain:
	\begin{equation}
		\label{33}
		X =-\xi , \quad Y = \pm \sqrt{\eta}.
	\end{equation}
	
		\begin{figure*}[htbp]
		\centering
		\begin{subfigure}{0.45\textwidth}
			\includegraphics[width=3in, height=5.5in, keepaspectratio]{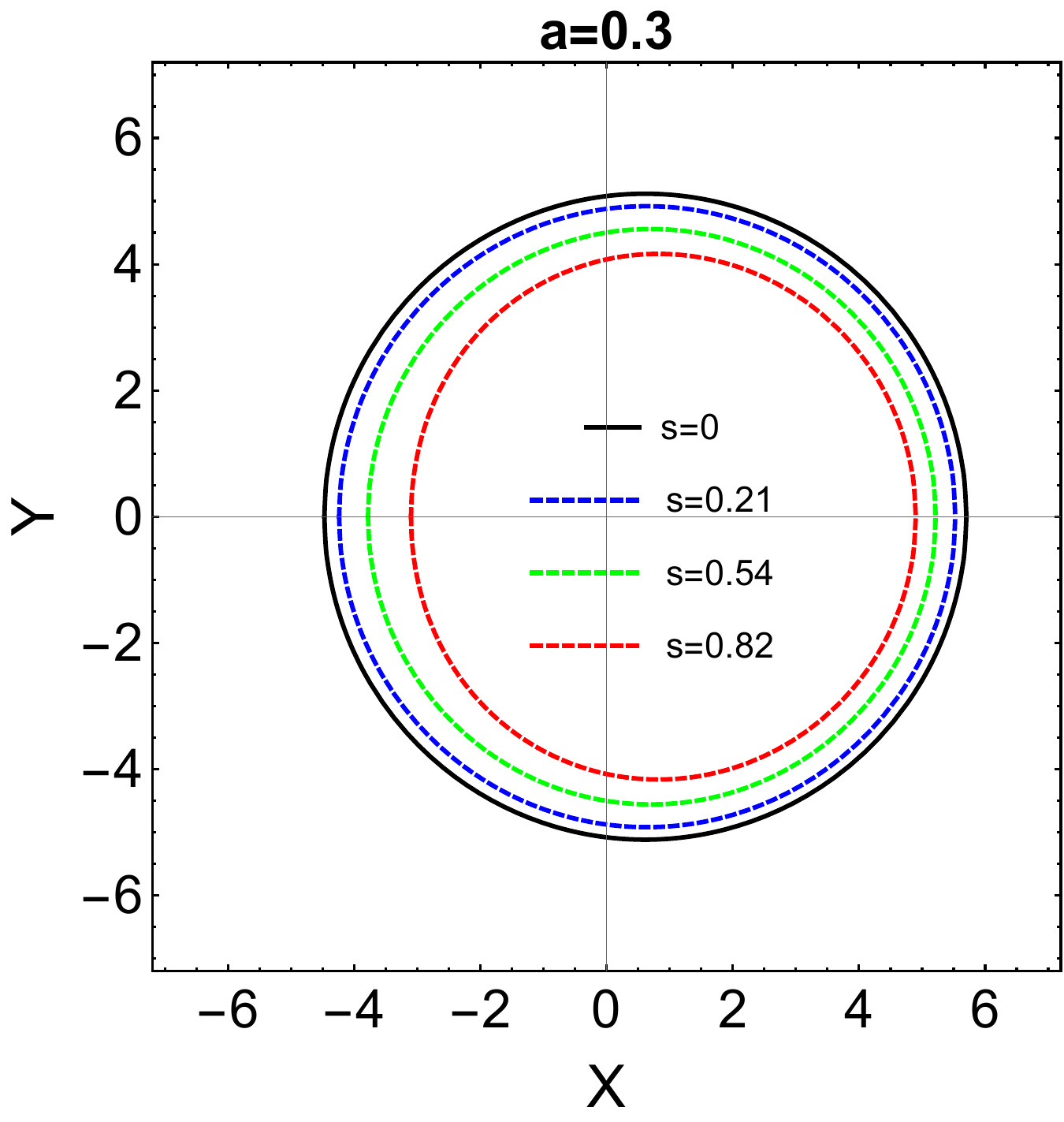}
		\end{subfigure}
		\hfill
		\begin{subfigure}{0.45\textwidth}
			\includegraphics[width=3in, height=5.5in,keepaspectratio]{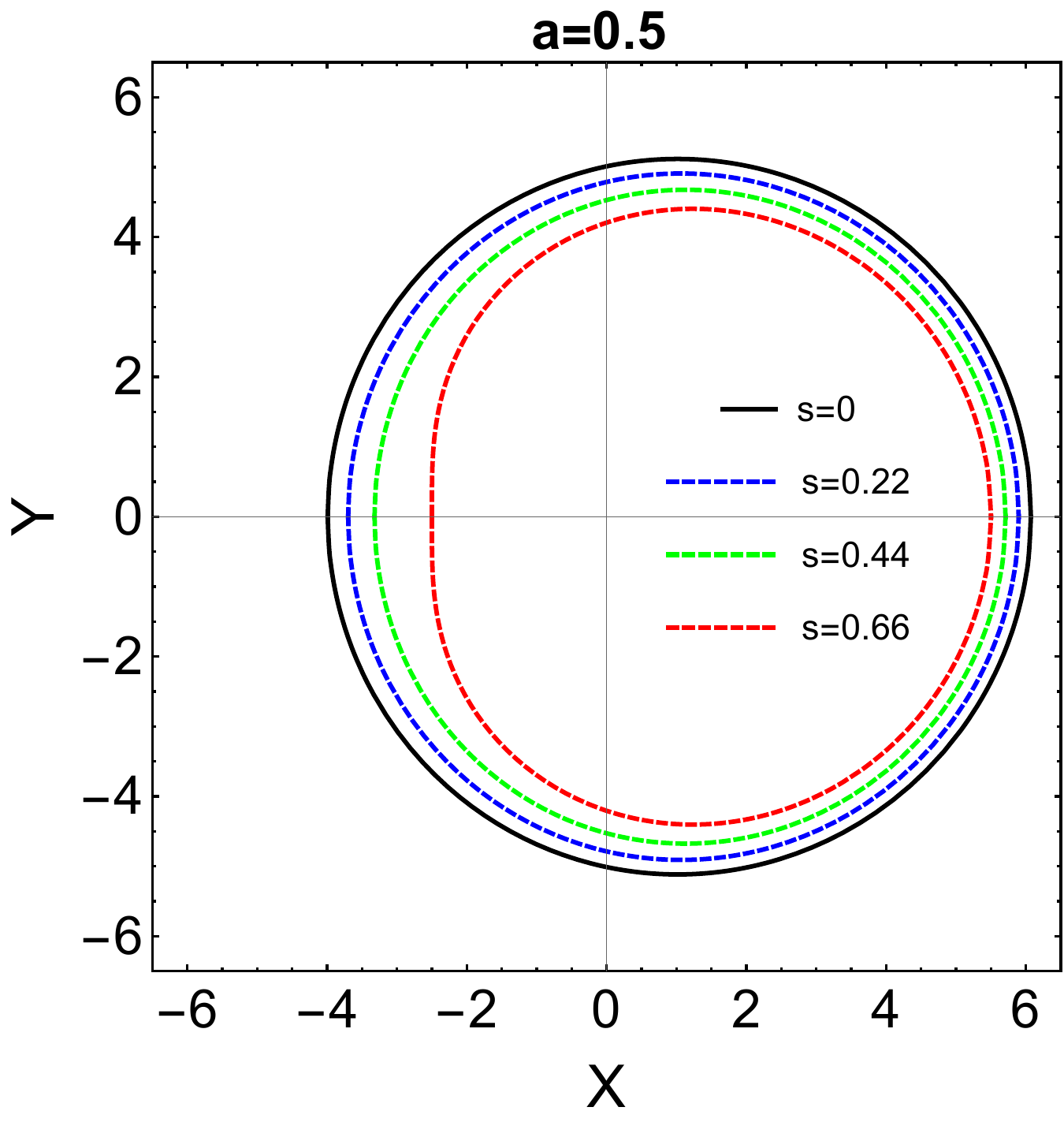}
		\end{subfigure}
		\begin{subfigure}{0.45\textwidth}
			\includegraphics[width=3in, height=5.5in, keepaspectratio]{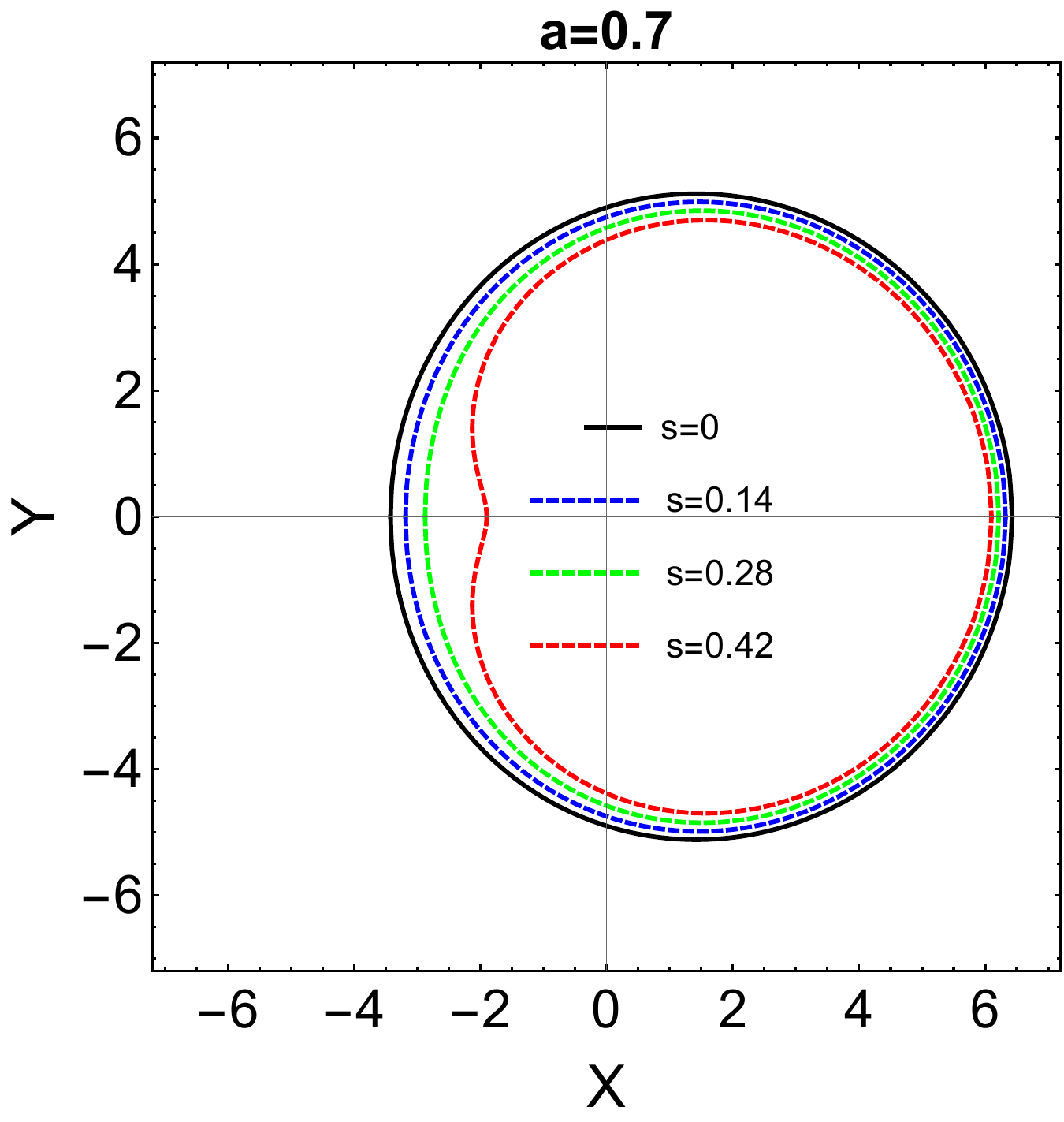}
		\end{subfigure}
		\hfill
		\begin{subfigure}{0.45\textwidth}
			\includegraphics[width=3in, height=5.5in, keepaspectratio]{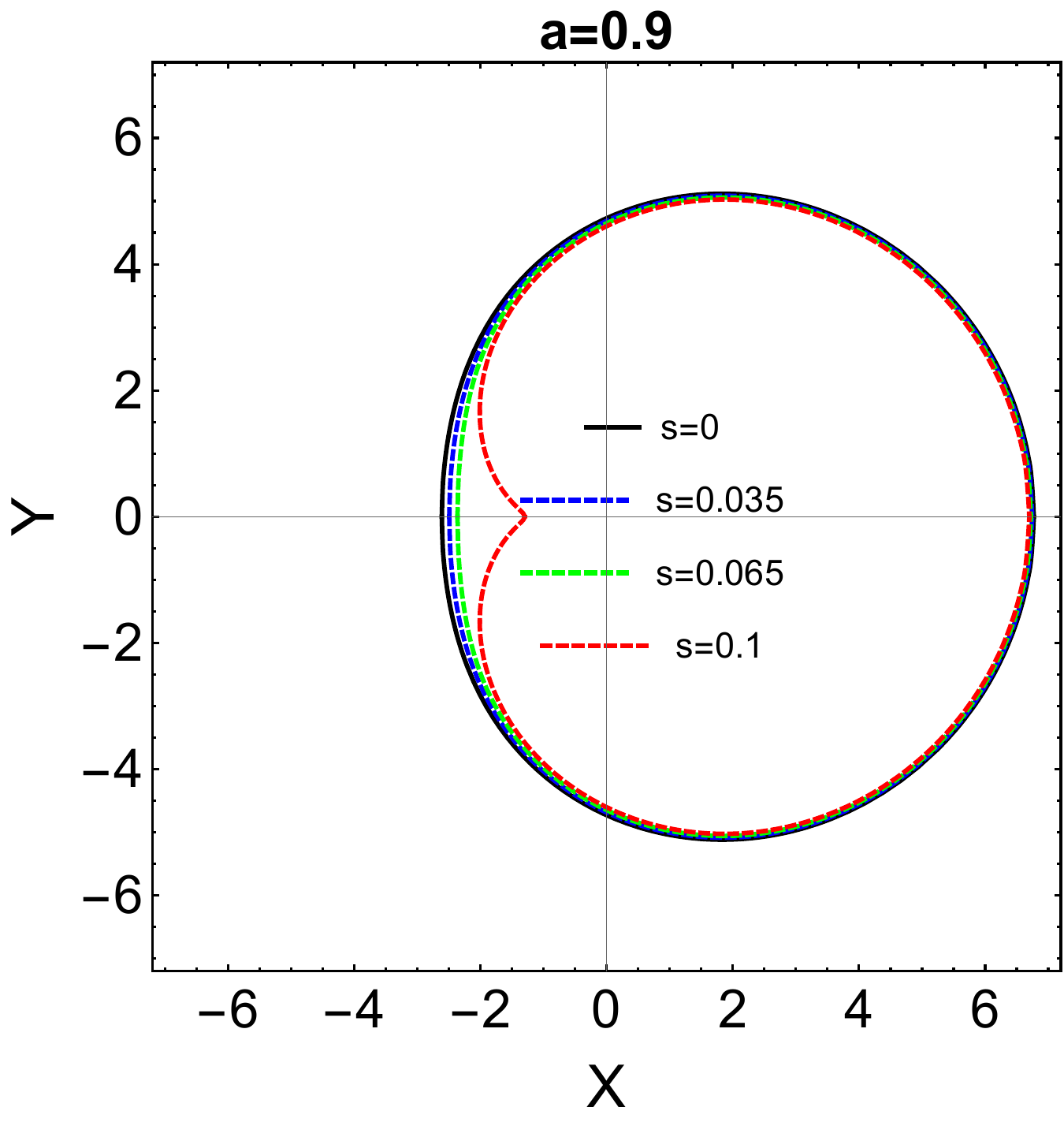}
		\end{subfigure}
		\caption{Shadows of a rotating charged black hole with scalar hair are presented for varying spin parameter $a$ and scalar hair parameter $s$, with $Q = 0.3$ fixed.}
		\label{shadow1}
	\end{figure*}
	
	\begin{figure*}[htbp]
		\centering
		\begin{subfigure}{0.45\textwidth}
			\includegraphics[width=3in, height=5.5in, keepaspectratio]{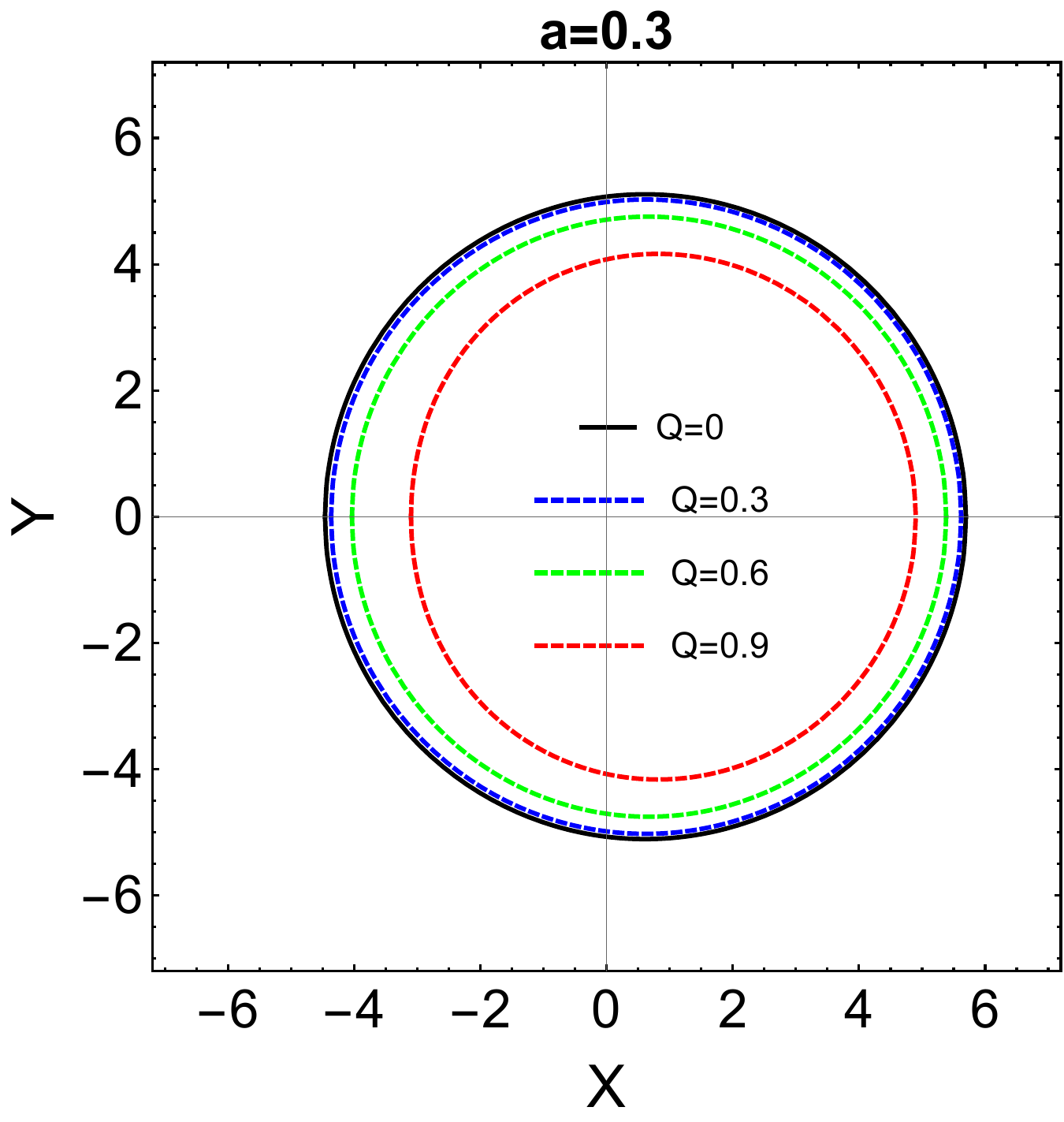}
		\end{subfigure}
		\hfill
		\begin{subfigure}{0.45\textwidth}
			\includegraphics[width=3in, height=5.5in,keepaspectratio]{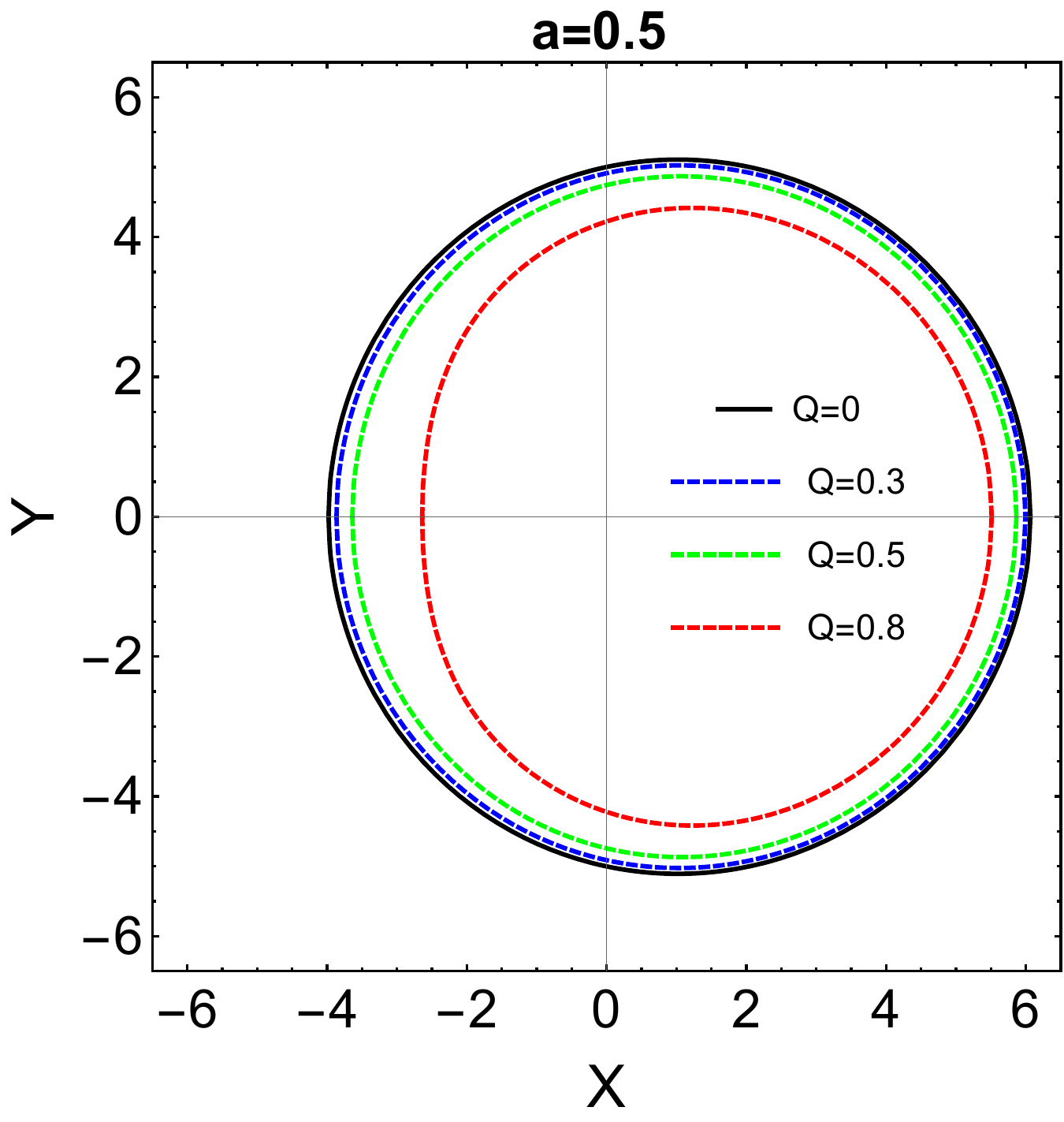}
		\end{subfigure}
		\begin{subfigure}{0.45\textwidth}
			\includegraphics[width=3in, height=5.5in, keepaspectratio]{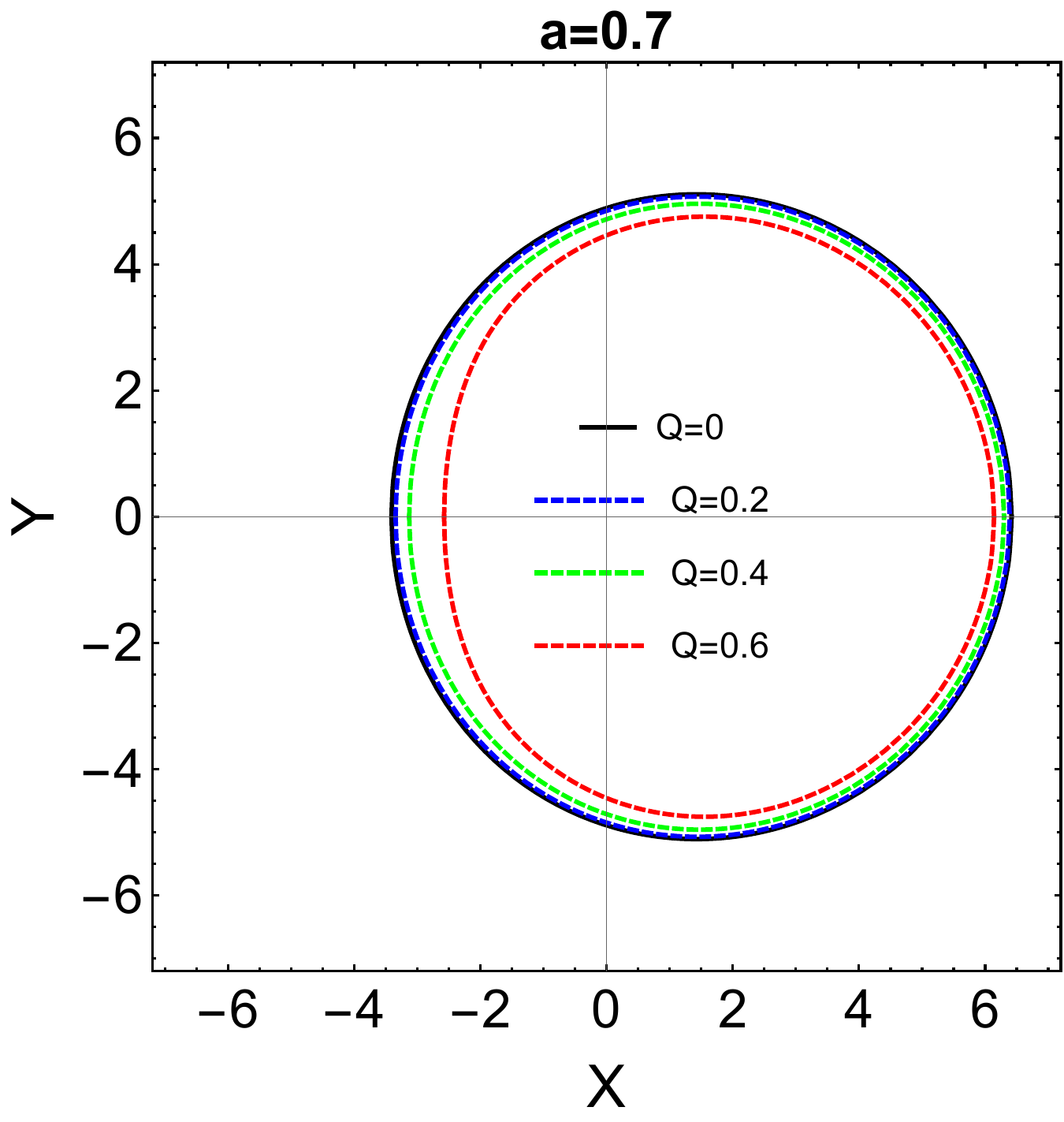}
		\end{subfigure}
		\hfill
		\begin{subfigure}{0.45\textwidth}
			\includegraphics[width=3in, height=5.5in, keepaspectratio]{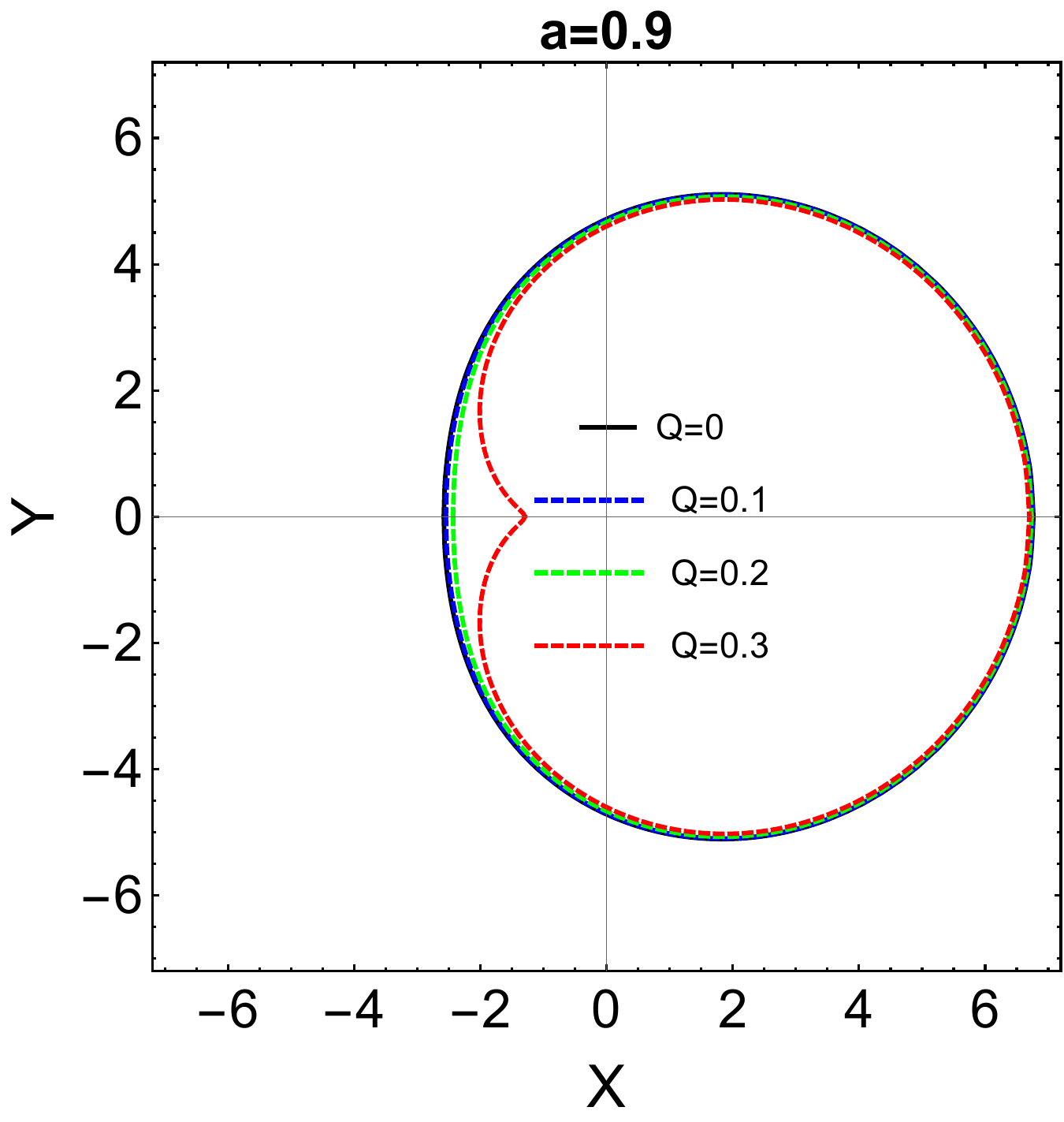}
		\end{subfigure}
		\caption{Shadows of a rotating charged black hole with scalar hair are shown for varying spin parameter $a$ and charge parameter $Q$, with $s = 0.1$ fixed. }
		\label{shadow2}
	\end{figure*}
	
	Figures~\ref{shadow1} and \ref{shadow2} show two-dimensional profiles of the shadow cast by a rotating charged BH with scalar hair. Figure~\ref{shadow1} focuses on the effect of the scalar hair parameter $s$ on the shadow, with the charge parameter fixed at $Q = 0.3$. For fixed spin $a$, increasing $s$ from zero leads to a monotonic decrease in the apparent size of the shadow (i.e., the shadow boundary contracts toward the center). At high spin values (e.g., $a = 0.9$), larger $s$ not only shrinks the shadow but also induces a distinctive cuspy edge, markedly different from the nearly circular shadows observed at low spin (e.g., $a = 0.3$). A cuspy edge also emerges at $a = 0.7$ as $s$ increases, but the effect is more pronounced for $a = 0.9$. In contrast, Figure~\ref{shadow2} investigates the influence of the charge parameter $Q$, with the scalar hair parameter fixed at $s = 0.1$. Here, for fixed $a$, increasing $Q$ similarly reduces the shadow size monotonically, and the deviation from the uncharged case becomes increasingly evident. 
	At high spins (e.g., $a = 0.9$), larger $Q$ enhances the non-circular distortion and also gives rise to a cuspy edge. A common trend emerges from both figures: regardless of whether $Q$ or $s$ is held fixed, increasing the other parameter (i.e., $s$ or $Q$) reduces the shadow size for a given $a$, and cuspy edges are more readily formed at high spin. However, a key difference exists: for the same $a$, the scalar hair parameter $s$ has a stronger effect on shadow deformation than the charge parameter $Q$. For instance, at $a = 0.7$, a cuspy edge appears when $s = 0.42$, whereas no such feature is observed for $Q = 0.6$, indicating that scalar hair induces more significant shape distortions than electric charge.
	
	Let us now shift our focus to the actual size and distortion of the shadow cast by a rotating charged BH with scalar hair. We introduce two observables that serve to characterize the BH shadow: $R_{s}$ and $\delta_{s}$. Specifically, the parameter $R_{s}$ denotes the actual size of the shadow, whereas $\delta_{s}$ represents the distortion in the shadow's shape. These parameters will allow for a more detailed analysis of the BH's shadow within the context of its observable characteristics. To determine the radius $R_{s}$, we use three points on the shadow boundary: the top point ($x_{t}, y_{t}$), the bottom point ($x_{b}, y_{b}$), and the rightmost point ($x_{r}, 0$). As defined in \cite{Hioki:2009na}:   
	\begin{equation}
		R_{s}=\frac{(x_{r}-x_{t})^{2}+y_{t}^{2}}{2(x_{r}-x_{t})}.\label{34}
	\end{equation} 
	The deformation $\delta_{s}$, describing how the shadow's shape deviates from a standard circle, is defined in \cite{Hioki:2009na} as: 
	\begin{equation}
		\delta_{s}=\frac{D}{R_{s}}=\frac{|x_{l}-\tilde{x}_{l}|}{R_{s}}.\label{35}
	\end{equation}  
		\begin{figure*}[htbp]
		\centering
		\begin{subfigure}{0.45\textwidth}
			\includegraphics[width=3in, height=5.5in, keepaspectratio]{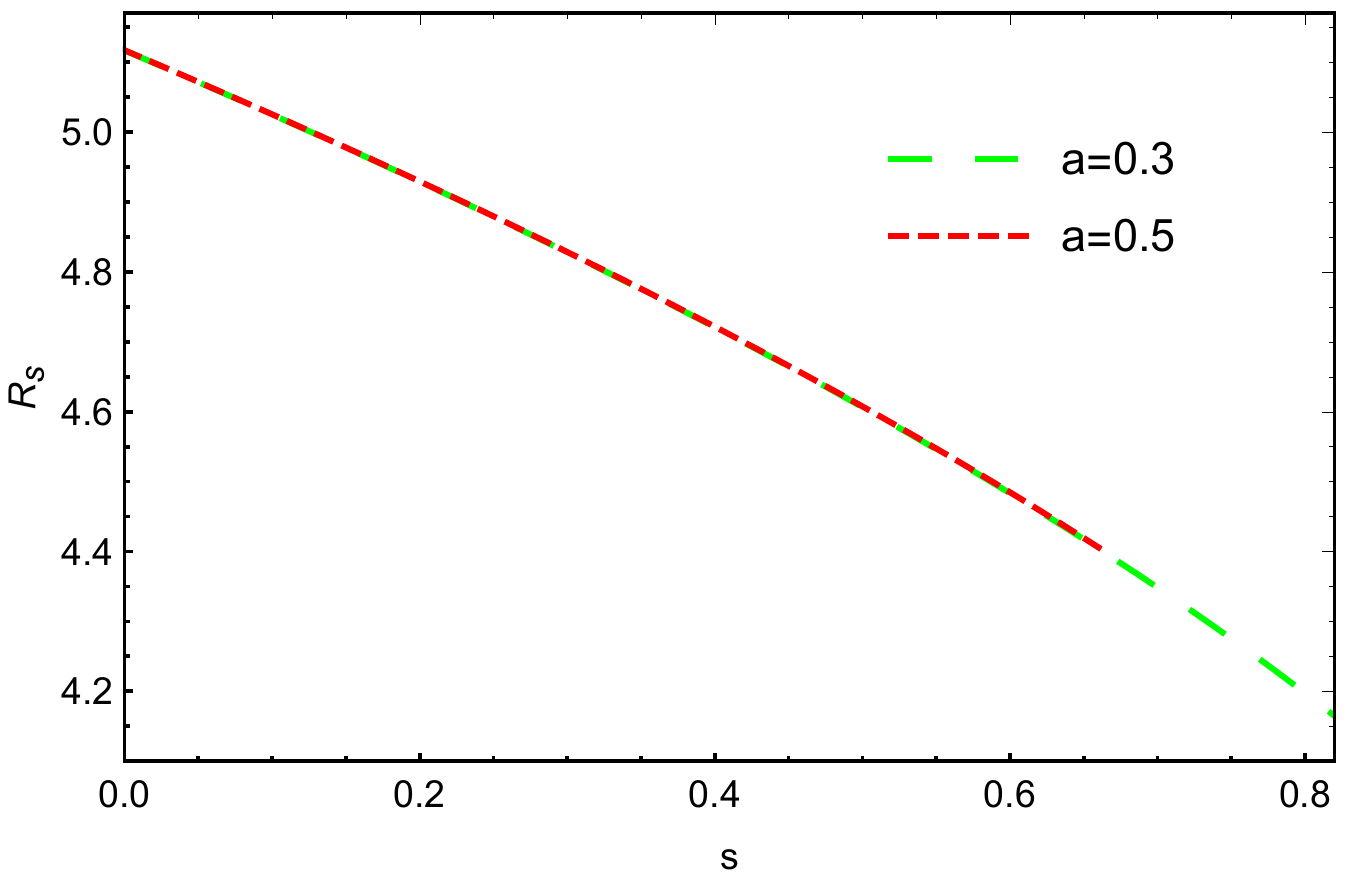}
		\end{subfigure}
		\hfill
		\begin{subfigure}{0.45\textwidth}
			\includegraphics[width=3in, height=5.5in,keepaspectratio]{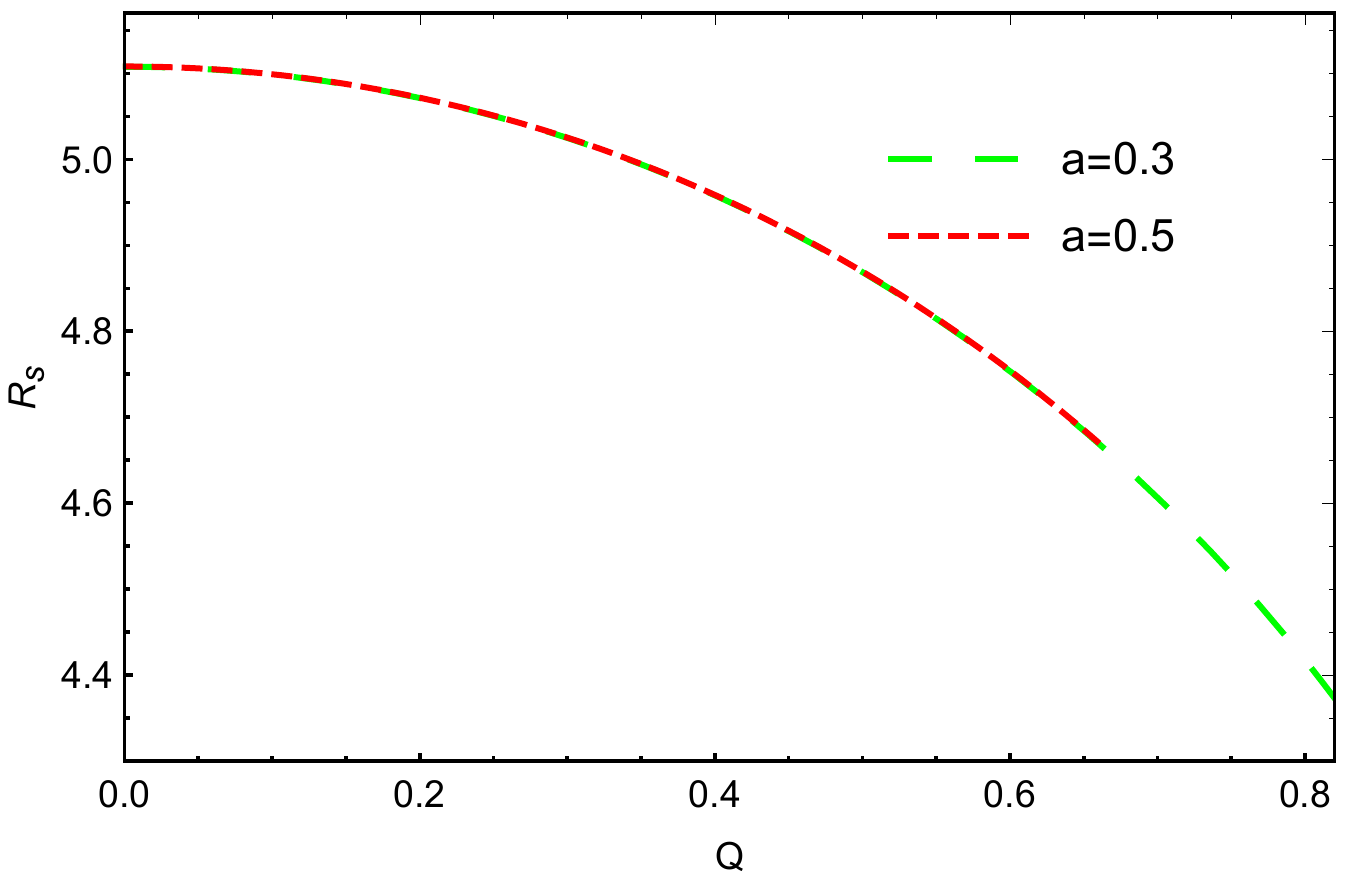}
		\end{subfigure}
		\begin{subfigure}{0.45\textwidth}
			\includegraphics[width=3in, height=5.5in, keepaspectratio]{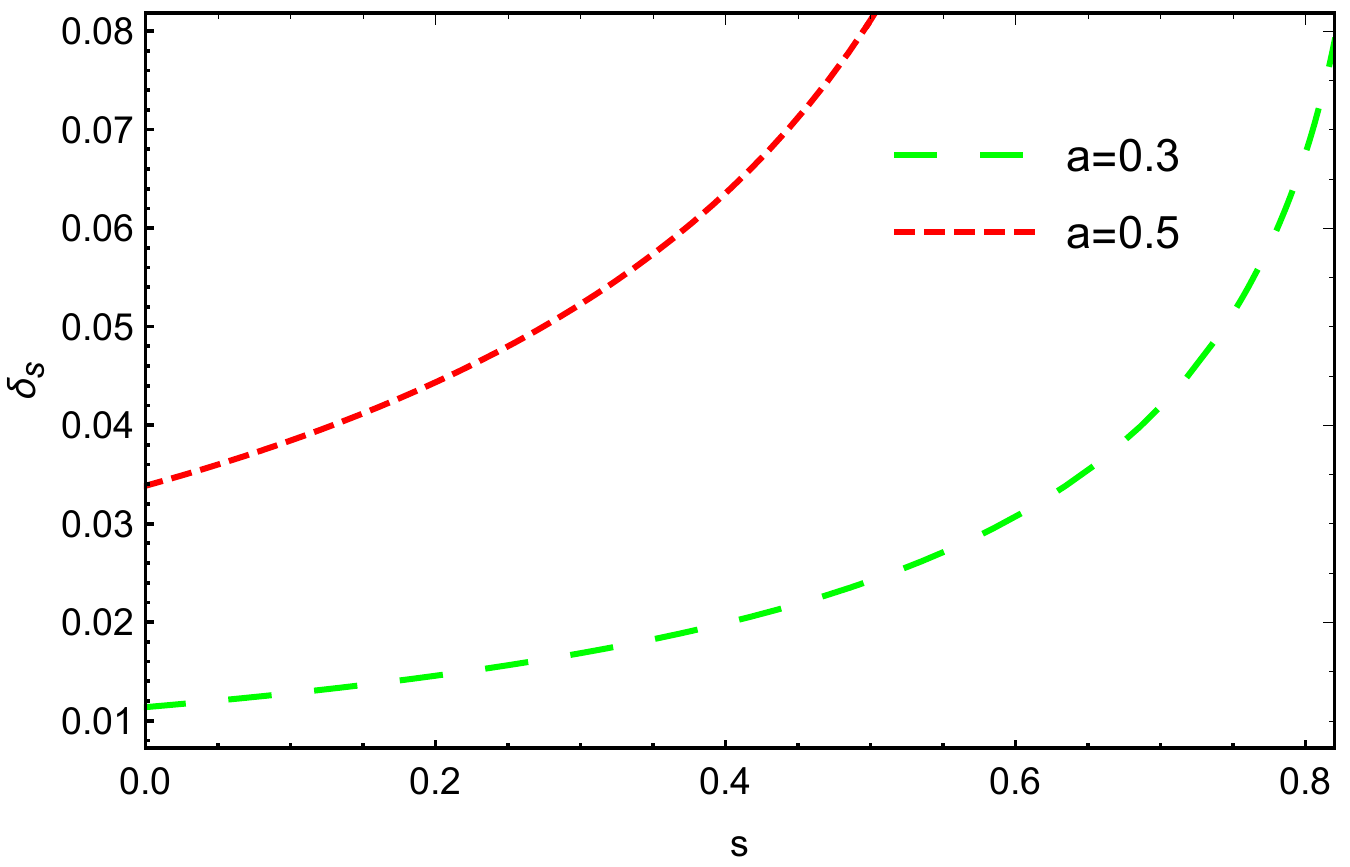}
		\end{subfigure}
		\hfill
		\begin{subfigure}{0.45\textwidth}
			\includegraphics[width=3in, height=5.5in, keepaspectratio]{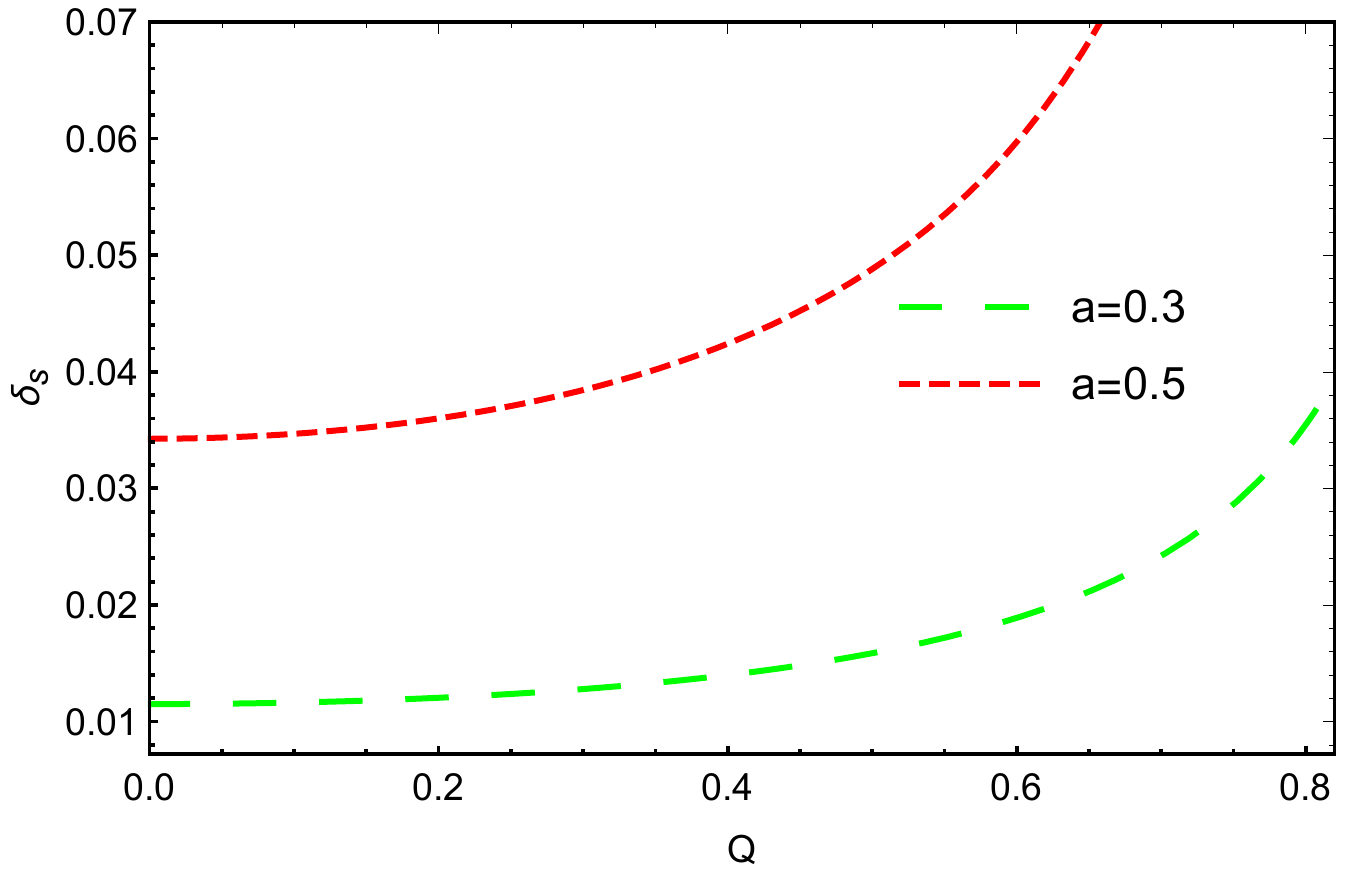}
		\end{subfigure}
		\caption{Variations of the shadow radius $R_s$ (top row) and deformation $\delta_s$ (bottom row) with scalar hair parameter $s$ (left column, $Q=0.3$) and charge $Q$ (right column, $s=0.1$), for spin parameters $a=0.3$ (green) and $a=0.5$ (red).}
		\label{Rdata}
	\end{figure*}
	Here, $D$ represents the distance between the leftmost point ($x_{l}, 0$) of the shadow and the rightmost point ($\tilde{x}_{l}, 0$) of the reference circle. Figure~\ref{Rdata} shows the variations of the shadow size $R_s$ and the shadow deformation $\delta_s$ as functions of the scalar hair parameter $s$ (left column) and the charge parameter $Q$ (right column), for BH spin parameters $a = 0.3$ and $a = 0.5$. In the top row, $R_s$ decreases monotonically with increasing $s$ (left panel, $Q = 0.3$ fixed) and with increasing $Q$ (right panel, $s = 0.1$ fixed), indicating that both scalar hair and electric charge reduce the apparent size of the shadow. Notably, the curves for $a = 0.3$ and $a = 0.5$ are nearly indistinguishable in both panels, implying that the variation of $R_s$ is insensitive to the spin parameter $a$ within this parameter range. In the bottom row, $\delta_s$ increases with both $s$ (left panel, $Q = 0.3$) and $Q$ (right panel, $s = 0.1$), reflecting growing deviations from circularity as either parameter increases. In contrast to $R_s$, the deformation $\delta_s$ depends strongly on $a$: for both $s$ and $Q$ scans, the $\delta_s$ curve for $a = 0.5$ lies systematically above that for $a = 0.3$, demonstrating that higher spin enhances the shadow's distortion. This indicates that while $R_s$ is primarily governed by $s$ and $Q$, $\delta_s$ is significantly influenced by the BH's rotation.
	
	\section{CONSTRAINTS FROM THE EHT OBSERVATIONS OF SUPERMASSIVE BLACK HOLES OF SGR A$^{*}$}\label{section5}
	The EHT observations of Sgr A$^{*}$ impose significant constraints on theoretical models of BHs. By imaging the BH shadow and its surrounding emission, the EHT data enables precise constraints on fundamental parameters. These observations provide unprecedented insights into the nature of BHs and their immediate environments. In this section, we use these data to constrain the parameter space of rotating charged BHs with scalar hair.
	
\begin{figure*}[htbp]
		\begin{minipage}[t]{0.45\textwidth}
			\centering
			\begin{overpic}[width=0.8\textwidth]{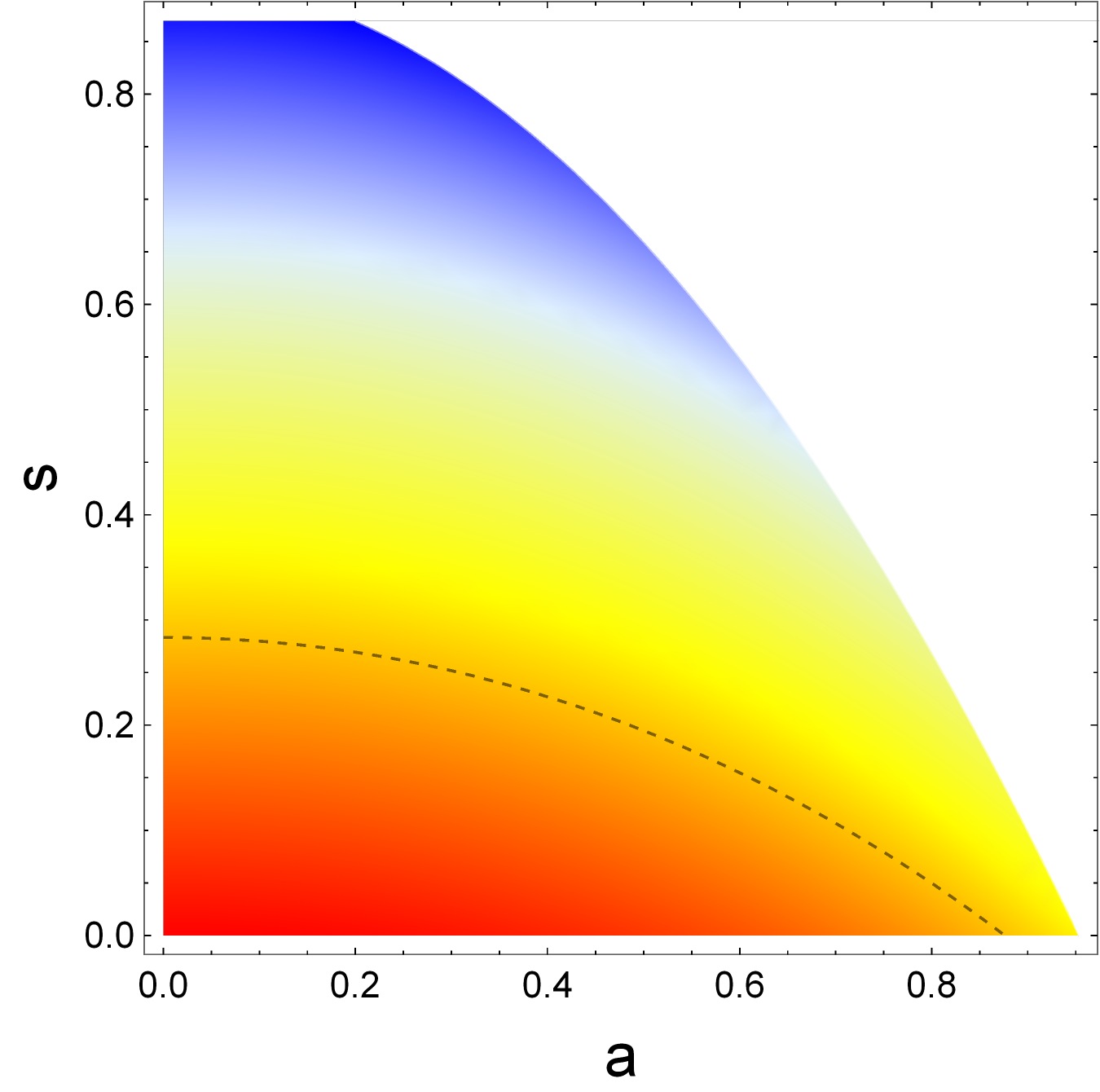} 
				\put(45,100){\color{black}\large $\theta_0=17^\circ$} 
			\end{overpic}
			\raisebox{0.087\height}{
				\begin{overpic}[width=0.097\textwidth]{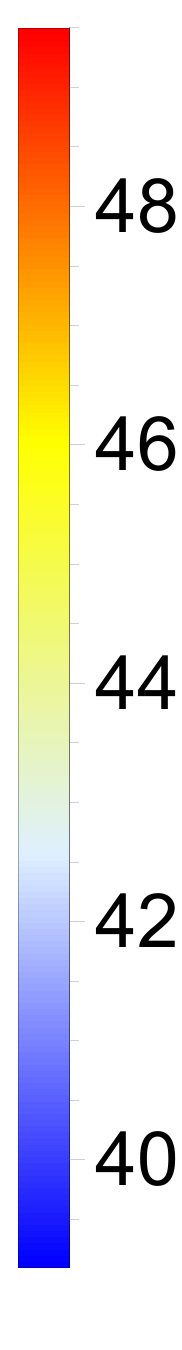}  
				\end{overpic}
			}
		\end{minipage}
		\hfill 
		\begin{minipage}[t]{0.45\textwidth}
			\centering
			\begin{overpic}[width=0.8\textwidth]{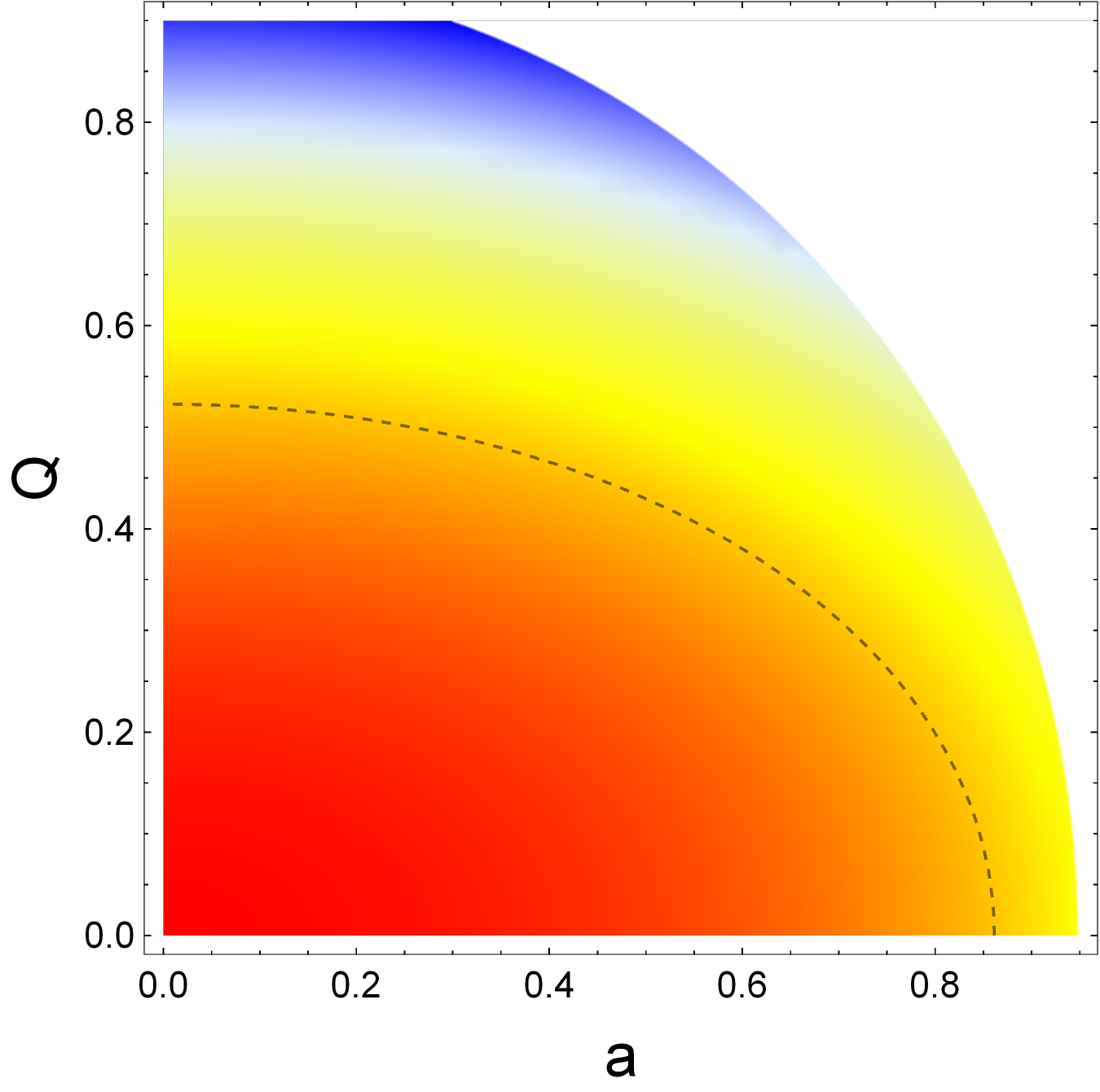} 
				\put(45,100){\color{black}\large $\theta_0=17^\circ$}
			\end{overpic}
			\raisebox{0.087\height}{
				\begin{overpic}[width=0.097\textwidth]{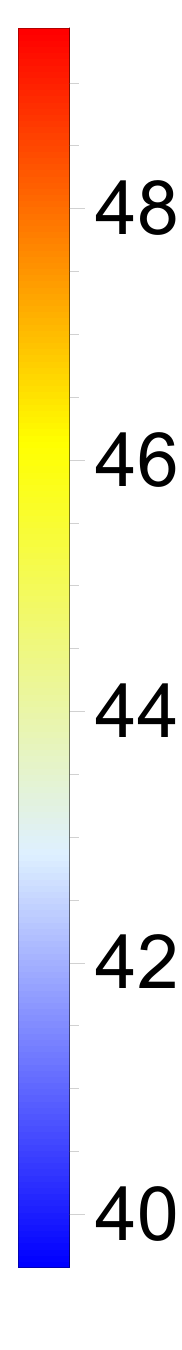}
				\end{overpic}
			}
		\end{minipage}
		\caption{Density plots of the shadow angular diameter $\theta_d$ for rotating charged black holes with scalar hair at a viewing angle $\theta_0 = 17^\circ$. The left panel fixes the charge parameter $Q = 0.3$, while the right panel fixes the scalar hair parameter $s = 0.1$. The dashed black line indicates $\theta_d = 46.9\,\mu\mathrm{as}$, corresponding to the lower limit of the angular diameter inferred from EHT observations of Sgr A$^{*}$.}
		\label{SgrA1}
	\end{figure*}

\begin{figure*}[htbp]
		\begin{minipage}[t]{0.45\textwidth}
			\centering
			\begin{overpic}[width=0.8\textwidth]{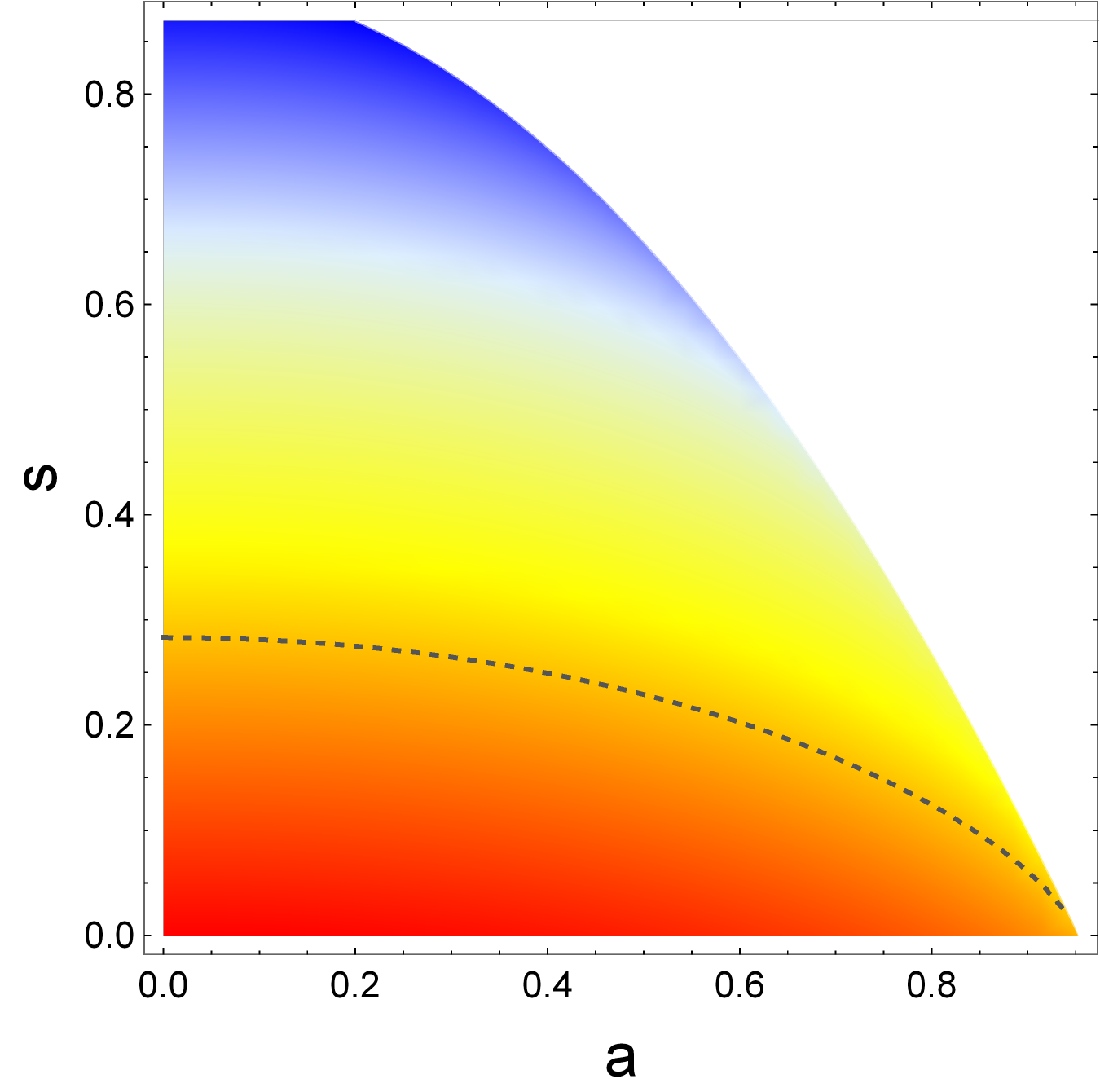}
				\put(45,100){\color{black}\large $\theta_0=90^\circ$}
			\end{overpic}
			\raisebox{0.087\height}{
				\begin{overpic}[width=0.097\textwidth]{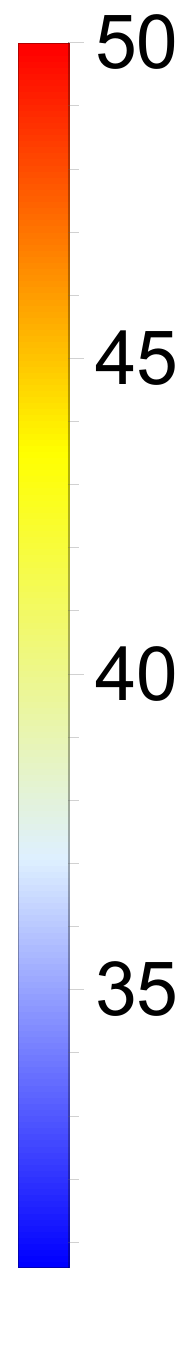}
				\end{overpic}
			}
		\end{minipage}%
		\hfill 
		\begin{minipage}[t]{0.45\textwidth}
			\centering
			\begin{overpic}[width=0.8\textwidth]{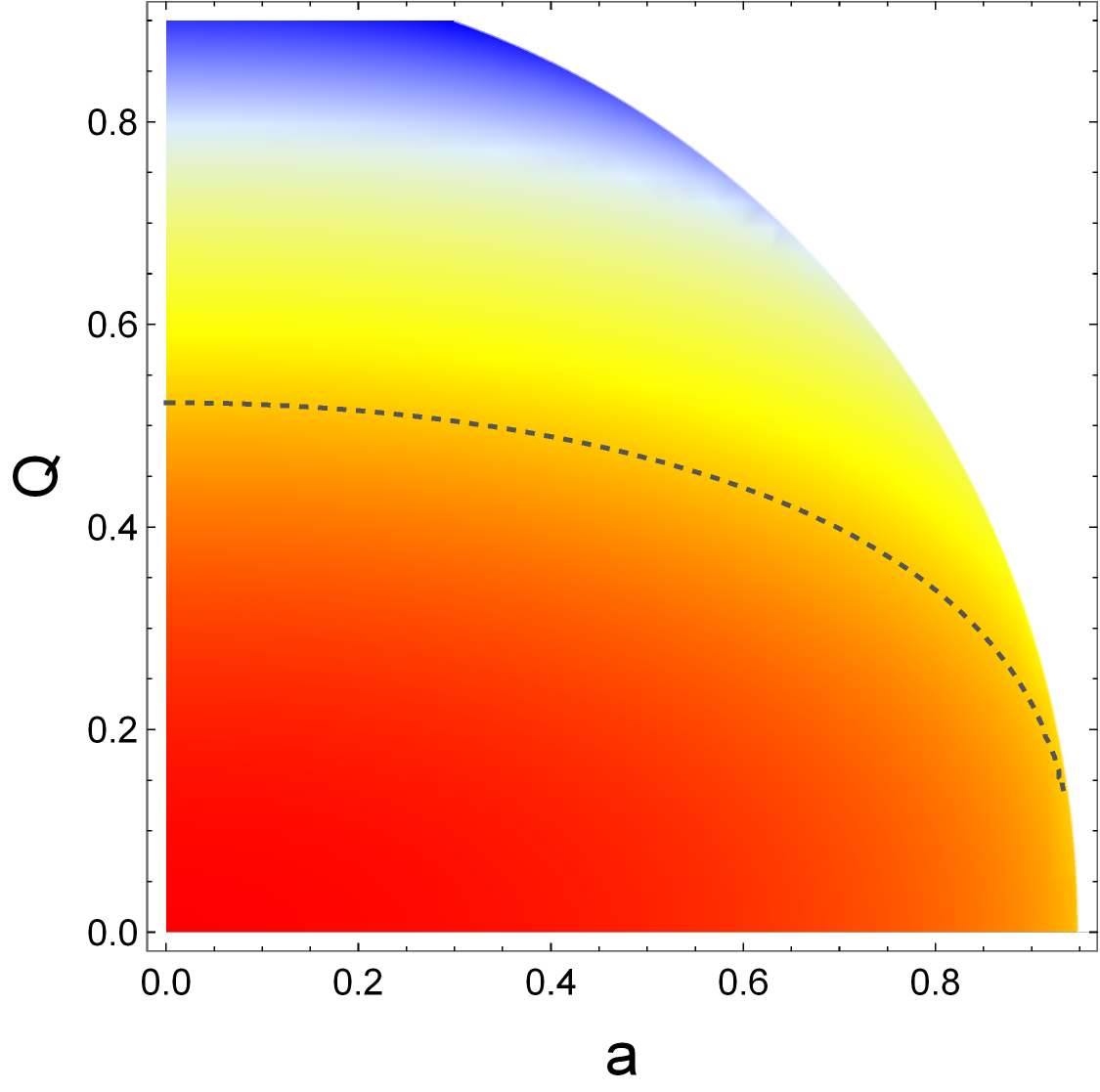}
				\put(45,100){\color{black}\large $\theta_0=90^\circ$}
			\end{overpic}
			\raisebox{0.087\height}{
				\begin{overpic}[width=0.097\textwidth]{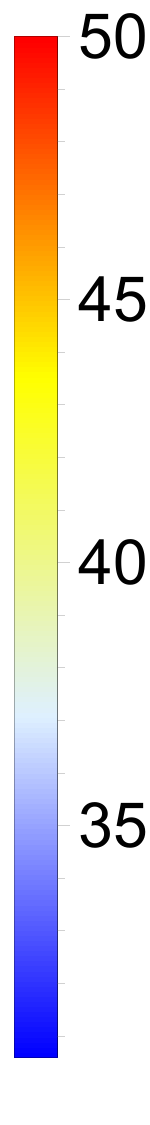} 
				\end{overpic}
			}
		\end{minipage}
		\caption{Density plots of the shadow angular diameter $\theta_d$ for rotating charged black holes with scalar hair at a viewing angle $\theta_0 = 90^\circ$. The left panel fixes the charge parameter $Q = 0.3$, while the right panel fixes the scalar hair parameter $s = 0.1$. The dashed black line indicates $\theta_d = 46.9\,\mu\mathrm{as}$, corresponding to the lower limit of the angular diameter inferred from EHT observations of Sgr A$^{*}$.}
		\label{SgrA2}
	\end{figure*}

	An areal radius $R_{s}=\sqrt{A/\pi}$ defines the BH shadow, which corresponds to an angular diameter $\theta_{d}$ as seen from the observer's celestial sphere:
	\begin{equation}
		\theta_{d}=2\frac{R_{s}}{d}=\frac{2}{d}\sqrt{\frac{A}{\pi}},\label{39}
	\end{equation} 
	here, $A$ is the BH shadow area, and $d$ represents the distance between the BH and Earth. Based on stellar-dynamical observations of the S0-2 star's orbit, conducted with the Keck Telescopes and the Very Large Telescope Interferometer (VLTI), the mass of Sgr A$^{*}$	is determined to be $M=4.0_{-0.6}^{+1.1}\times10^{6}M_{\odot}$, at a distance of $d=8$ kpc. The EHT image of Sgr A$^{*}$ shows an angular shadow diameter within the range $\theta_{d}\in(46.9, 50)\mu$as, which is consistent with the expected shadow of a Kerr BH. Figures~\ref{SgrA1} and \ref{SgrA2} display the shadow angular diameter of rotating charged BHs with scalar hair at inclination angles $\theta_0 = 17^\circ$ and $90^\circ$, respectively. The figure consists of two panels: the left panel fixes the charge parameter at $Q = 0.3$, while the right panel fixes the scalar hair parameter at $s = 0.1$. The dashed black line marks the lower limit of the Sgr A$^{*}$ shadow size, $46.9\,\mu$as. At $\theta_0 = 17^\circ$, we derive upper bounds on $a$, yielding $a < 0.874$ (left) and $a < 0.861$ (right), along with constraints on $s < 0.283373$ and $Q < 0.522745$. At $\theta_0 = 90^\circ$, the model predictions exceed the observational lower bound for all $a$, so no constraint on the spin arises. However, $s < 0.283373$ and $Q < 0.522745$ still hold.
	
	\section{Conclusion}\label{section6}
	
    In this study, we have derived a rotating charged BH solution with scalar hair by applying the NJA to the spherically symmetric charged BH with scalar hair. We have systematically investigated the physical properties of the event horizon and ergosurface in this spacetime. Our analysis shows that the number and structure of horizons depend on the interplay between the charge parameter $Q$, the scalar hair parameter $s$, and the spin parameter $a$. Notably, for fixed $a$ and $Q$, both the radius of the outer horizon and the outer ergosphere decrease as $s$ increases. Similarly, for fixed $a$ and $s$, increasing the charge $Q$ leads to a reduction in both the outer horizon and ergosphere radii.
	
	Additionally, we have applied the Novikov-Thorne model to study thin accretion disks around rotating charged BHs with scalar hair. 
	For fixed spin parameter $a$, the energy flux and radiation temperature of the disk exhibit qualitatively similar radial profiles: they rise with radius, reach a maximum, and then decline. 
	Both quantities increase with either the charge parameter $Q$ or the scalar hair parameter $s$, indicating enhanced radiative efficiency in the presence of larger $Q$ or $s$. 
	Conversely, for fixed $Q$ or $s$, the peak values of the energy flux and temperature decrease as $a$ increases, suggesting that higher spin reduces the peak emission in this class of BH models with scalar hair.
	
	Furthermore, we have analyzed how the parameters $a$, $Q$, and $s$ influence the shape and size of the shadow cast by rotating charged BHs with scalar hair. Our results show that, for fixed spin $a$, the shadow angular diameter decreases as either the charge $Q$ or the scalar hair parameter $s$ increases. In near-extremal cases, the shadow develops a cuspy edge, reflecting strong light bending in the vicinity of the horizon. Comparing with EHT observations of Sgr A$^{*}$, we obtain the constraints $0<Q<0.522745$ (at $s = 0.1$) and $0<s<0.283373$ (at $Q = 0.3$). These results demonstrate that rotating charged BHs with scalar hair are compatible with current BH imaging data.

	%-------------------------------------------------------------
	\begin{acknowledgments}
		This research was supported by the National Natural Science Foundation of China (Grant No. 12265007).
	\end{acknowledgments}
	
	%===============================================================================================
	%-------------------
	
	%===============================================================================================
	%===============================================================================================
	
	%===============================================================================================
	

\begin{thebibliography}{80}
		
		\bibitem{EventHorizonTelescope:2019dse}
		K.~Akiyama \textit{et al.} [Event Horizon Telescope],
		{First M87 Event Horizon Telescope Results. I. The Shadow of the Supermassive Black Hole},
		\href{http://dx.doi.org/10.3847/2041-8213/ab0ec7}{Astrophys. J. Lett. {\bfseries 875}, L1 (2019)},
		[\href{http://arxiv.org/abs/1906.11238}{arXiv:1906.11238 [astro-ph.GA]}].
		
		\bibitem{EventHorizonTelescope:2019uob}
		K.~Akiyama \textit{et al.} [Event Horizon Telescope],
		{First M87 Event Horizon Telescope Results. II. Array and Instrumentation},
		\href{http://dx.doi.org/10.3847/2041-8213/ab0c96}{Astrophys. J. Lett. {\bfseries 875}, no.1, L2 (2019)},
		[\href{http://arxiv.org/abs/1906.11239}{arXiv:1906.11239 [astro-ph.IM]}].
		
		\bibitem{EventHorizonTelescope:2019jan}
		K.~Akiyama \textit{et al.} [Event Horizon Telescope],
		{First M87 Event Horizon Telescope Results. III. Data Processing and Calibration},
		\href{http://dx.doi.org/10.3847/2041-8213/ab0c57}{Astrophys. J. Lett. {\bfseries 875}, no.1, L3 (2019)},
		[\href{http://arxiv.org/abs/1906.11240}{arXiv:1906.11240 [astro-ph.GA]}].
		
		\bibitem{EventHorizonTelescope:2019ths}
		K.~Akiyama \textit{et al.} [Event Horizon Telescope],
		{First M87 Event Horizon Telescope Results. IV. Imaging the Central Supermassive Black Hole},
		\href{http://dx.doi.org/10.3847/2041-8213/ab0e85}{Astrophys. J. Lett. {\bfseries 875}, no.1, L4 (2019)},
		[\href{http://arxiv.org/abs/1906.11241}{arXiv:1906.11241 [astro-ph.GA]}].
		
		\bibitem{EventHorizonTelescope:2019pgp}
		K.~Akiyama \textit{et al.} [Event Horizon Telescope],
		{First M87 Event Horizon Telescope Results. V. Physical Origin of the Asymmetric Ring},
		\href{http://dx.doi.org/10.3847/2041-8213/ab0f43}{Astrophys. J. Lett. {\bfseries 875}, no.1, L5 (2019)},
		[\href{http://arxiv.org/abs/1906.11242}{arXiv:1906.11242 [astro-ph.GA]}].
		
		\bibitem{EventHorizonTelescope:2019ggy}
		K.~Akiyama \textit{et al.} [Event Horizon Telescope],
		{First M87 Event Horizon Telescope Results. VI. The Shadow and Mass of the Central Black Hole},
		\href{http://dx.doi.org/10.3847/2041-8213/ab1141}{Astrophys. J. Lett. {\bfseries 875}, no.1, L6 (2019)},
		[\href{http://arxiv.org/abs/1906.11243}{arXiv:1906.11243 [astro-ph.GA]}].	
		
		\bibitem{EventHorizonTelescope:2022wkp}
		K.~Akiyama \textit{et al.} [Event Horizon Telescope],
		{First Sagittarius A* Event Horizon Telescope Results. I. The Shadow of the Supermassive Black Hole in the Center of the Milky Way},
		\href{http://dx.doi.org/10.3847/2041-8213/ac6674}{Astrophys. J. Lett. {\bfseries 930}, no.2, L12 (2022)},
		[\href{http://arxiv.org/abs/2311.08680}{arXiv:2311.08680 [astro-ph.HE]}].
		
		\bibitem{Shakura:1972te}
		N.~I.~Shakura and R.~A.~Sunyaev,
		{Black holes in binary systems. Observational appearance},
		\href{http://dx.doi.org/10.1051/0004-6361:19730038}{Astron. Astrophys. {\bfseries 24}, 337-355 (1973)},
		[\href{http://arxiv.org/abs/astro-ph/0301586}{astro-ph/0301586}].
		
		\bibitem{Page:1974he}
		D.~N.~Page and K.~S.~Thorne,
		{Disk-Accretion onto a Black Hole. Time-Averaged Structure of Accretion Disk},
		\href{http://dx.doi.org/10.1086/152990}{Astrophys. J. {\bfseries 191}, 499-506 (1974)},
		[\href{http://arxiv.org/abs/astro-ph/0301586}{astro-ph/0301586}].
		
		\bibitem{Harko:2009rp}
		T.~Harko, Z.~Kovacs and F.~S.~N.~Lobo,
		{Testing Ho\v{r}ava-Lifshitz gravity using thin accretion disk properties},
		\href{http://dx.doi.org/10.1103/PhysRevD.80.044021}{Phys. Rev. D {\bfseries 80}, no.4, 044021 (2009)},
		[\href{http://arxiv.org/abs/0907.1449}{arXiv:0907.1449 [gr-qc]}].
		
		\bibitem{Chen:2011wb}
		S.~Chen and J.~Jing,
		{Thin accretion disk around a Kaluza\textendash{}Klein black hole with squashed horizons},
		\href{http://dx.doi.org/10.1016/j.physletb.2011.09.071}{Phys. Lett. B {\bfseries 704}, 641-645 (2011)},
		[\href{http://arxiv.org/abs/1106.5183}{arXiv:1106.5183 [gr-qc]}].
		
		\bibitem{Liu:2021yev}
		C.~Liu, S.~Yang, Q.~Wu and T.~Zhu,
		{Thin accretion disk onto slowly rotating black holes in Einstein-\AE{}ther theory},
		\href{http://dx.doi.org/10.1088/1475-7516/2022/02/034}{JCAP {\bfseries 02}, no.02, 034 (2022)},
		[\href{http://arxiv.org/abs/2107.04811}{arXiv:2107.04811 [gr-qc]}].
		
		\bibitem{Heydari-Fard:2021ljh}
		M.~Heydari-Fard, M.~Heydari-Fard and H.~R.~Sepangi,
		{Thin accretion disks around rotating black holes in 4$D$ Einstein\textendash{}Gauss\textendash{}Bonnet gravity},
		\href{http://dx.doi.org/10.1140/epjc/s10052-021-09266-7}{Eur. Phys. J. C {\bfseries 81}, no.5, 473 (2021)},
		[\href{http://arxiv.org/abs/2105.09192}{arXiv:2105.09192 [gr-qc]}].
		
		\bibitem{Karimov:2018whx}
		R.~K.~Karimov, R.~N.~Izmailov, A.~Bhattacharya and K.~K.~Nandi,
		{Accretion disks around the Gibbons\textendash{}Maeda\textendash{}Garfinkle\textendash{}Horowitz\textendash{}Strominger charged black holes},
		\href{http://dx.doi.org/10.1140/epjc/s10052-018-6270-6}{Eur. Phys. J. C {\bfseries 78}, no.9, 788 (2018)},
		[\href{http://arxiv.org/abs/2002.00589}{arXiv:2002.00589 [gr-qc]}].
		
		\bibitem{Chen:2011rx}
		S.~Chen and J.~Jing,
		{Properties of a thin accretion disk around a rotating non-Kerr black hole},
		\href{http://dx.doi.org/10.1016/j.physletb.2012.03.047}{Phys. Lett. B {\bfseries 711}, 81-87 (2012)},
		[\href{http://arxiv.org/abs/1110.3462}{arXiv:1110.3462 [gr-qc]}].
		
		\bibitem{Kazempour:2022asl}
		S.~Kazempour, Y.~C.~Zou and A.~R.~Akbarieh,
		{Analysis of accretion disk around a black hole in dRGT massive gravity},
		\href{http://dx.doi.org/10.1140/epjc/s10052-022-10153-y}{Eur. Phys. J. C {\bfseries 82}, no.3, 190 (2022)},
		[\href{http://arxiv.org/abs/2203.05190}{arXiv:2203.05190 [gr-qc]}].
		
		\bibitem{Gyulchev:2019tvk}
		G.~Gyulchev, P.~Nedkova, T.~Vetsov and S.~Yazadjiev,
		{Image of the Janis-Newman-Winicour naked singularity with a thin accretion disk},
		\href{http://dx.doi.org/10.1103/PhysRevD.100.024055}{Phys. Rev. D {\bfseries 100}, no.2, 024055 (2019)},
		[\href{http://arxiv.org/abs/1905.05273}{arXiv:1905.05273 [gr-qc]}].
		
		\bibitem{Wu:2024sng}
		Y.~Wu, H.~Feng and W.~Q.~Chen,
		{Thin accretion disk around black hole in Einstein--Maxwell-scalar theory},
		\href{http://dx.doi.org/10.1140/epjc/s10052-024-13454-6}{Eur. Phys. J. C {\bfseries 84}, no.10, 1075 (2024)},
		[\href{http://arxiv.org/abs/2410.14113}{arXiv:2410.14113 [gr-qc]}].
		
		\bibitem{Feng:2024iqj}
		H.~Feng, R.~J.~Yang and W.~Q.~Chen,
		{Thin accretion disk and shadow of Kerr--Sen black hole in Einstein--Maxwell-dilaton--axion gravity},
		\href{http://dx.doi.org/10.1016/j.astropartphys.2024.103075}{Astropart. Phys. {\bfseries 166}, 103075 (2025)},
		[\href{http://arxiv.org/abs/2403.18541}{arXiv:2403.18541 [gr-qc]}].
		
		\bibitem{Liu:2024brf}
		A.~Liu, T.~Y.~He, M.~Liu, Z.~W.~Han and R.~J.~Yang,
		{Possible signatures of higher dimension in thin accretion disk around brane world black hole},
		\href{http://dx.doi.org/10.1088/1475-7516/2024/07/062}{JCAP {\bfseries 07}, 062 (2024)},
		[\href{http://arxiv.org/abs/2404.14131}{arXiv:2404.14131 [gr-qc]}].
		
		\bibitem{Bambi:2015kza}
		C.~Bambi,
		{Testing black hole candidates with electromagnetic radiation},
		\href{http://dx.doi.org/10.1103/RevModPhys.89.025001}{Rev. Mod. Phys. {\bfseries 89}, no.2, 025001 (2017)},
		[\href{http://arxiv.org/abs/1509.03884}{arXiv:1509.03884 [gr-qc]}].
		
		\bibitem{Synge:1966okc}
		J.~L.~Synge,
		{The Escape of Photons from Gravitationally Intense Stars},
		\href{http://dx.doi.org/10.1093/mnras/131.3.463}{Mon. Not. Roy. Astron. Soc. {\bfseries 131}, no.3, 463-466 (1966)},
		[\href{http://arxiv.org/abs/astro-ph/0301586}{astro-ph/0301586}].
		
		\bibitem{Luminet:1979nyg}
		J.~P.~Luminet,
		{Image of a spherical black hole with thin accretion disk},
		\href{http://dx.doi.org/10.1051/0004-6361/19790750228}{Astron. Astrophys. {\bfseries 75}, 228-235 (1979)},
		[\href{http://arxiv.org/abs/astro-ph/0301586}{astro-ph/0301586}].
		
		\bibitem{Bardeen:1973tla}
		J.~M.~Bardeen,
		{Timelike and null geodesics in the Kerr metric},
		in \emph{Proceedings, Ecole d'Eté de Physique Théorique: Les Astres Occlus : Les Houches, France, August, 1972},
		pp.~215-240 (1973).
		
		\bibitem{Stuchlik:2019uvf}
		Z.~Stuchl\'\i{}k and J.~Schee,
		{Shadow of the regular Bardeen black holes and comparison of the motion of photons and neutrinos},
		\href{http://dx.doi.org/10.1140/epjc/s10052-019-6543-8}{Eur. Phys. J. C {\bfseries 79}, no.1, 44 (2019)},
		[\href{http://arxiv.org/abs/1904.03809}{arXiv:1904.03809 [gr-qc]}].
		
		\bibitem{Yumoto:2012kz}
		A.~Yumoto, D.~Nitta, T.~Chiba and N.~Sugiyama,
		{Shadows of Multi-Black Holes: Analytic Exploration},
		\href{http://dx.doi.org/10.1103/PhysRevD.86.103001}{Phys. Rev. D {\bfseries 86}, 103001 (2012)},
		[\href{http://arxiv.org/abs/1208.0635}{arXiv:1208.0635 [gr-qc]}].
		
		\bibitem{Abdujabbarov:2016hnw}
		A.~Abdujabbarov, M.~Amir, B.~Ahmedov and S.~G.~Ghosh,
		{Shadow of rotating regular black holes},
		\href{http://dx.doi.org/10.1103/PhysRevD.93.104004}{Phys. Rev. D {\bfseries 93}, no.10, 104004 (2016)},
		[\href{http://arxiv.org/abs/1604.03809}{arXiv:1604.03809 [gr-qc]}].
		
		\bibitem{Amir:2016cen}
		M.~Amir and S.~G.~Ghosh,
		{Shapes of rotating nonsingular black hole shadows},
		\href{http://dx.doi.org/10.1103/PhysRevD.94.024054}{Phys. Rev. D {\bfseries 94}, no.2, 024054 (2016)},
		[\href{http://arxiv.org/abs/1603.06382}{arXiv:1603.06382 [gr-qc]}].
		
		\bibitem{Sharif:2016znp}
		M.~Sharif and S.~Iftikhar,
		{Shadow of a Charged Rotating Non-Commutative Black Hole},
		\href{http://dx.doi.org/10.1140/epjc/s10052-016-4472-3}{Eur. Phys. J. C {\bfseries 76}, no.11, 630 (2016)},
		[\href{http://arxiv.org/abs/1611.00611}{arXiv:1611.00611 [gr-qc]}].
		
		\bibitem{Wei:2013kza}
		S.~W.~Wei and Y.~X.~Liu,
		{Observing the shadow of Einstein-Maxwell-Dilaton-Axion black hole},
		\href{http://dx.doi.org/10.1088/1475-7516/2013/11/063}{JCAP {\bfseries 11}, 063 (2013)},
		[\href{http://arxiv.org/abs/1311.4251}{arXiv:1311.4251 [gr-qc]}].
		
		\bibitem{Abdujabbarov:2012bn}
		A.~Abdujabbarov, F.~Atamurotov, Y.~Kucukakca, B.~Ahmedov and U.~Camci,
		{Shadow of Kerr-Taub-NUT black hole},
		\href{http://dx.doi.org/10.1007/s10509-012-1337-6}{Astrophys. Space Sci. {\bfseries 344}, 429-435 (2013)},
		[\href{http://arxiv.org/abs/1212.4949}{arXiv:1212.4949 [physics.gen-ph]}].
		
		\bibitem{Neves:2020doc}
		J.~C.~S.~Neves,
		{Constraining the tidal charge of brane black holes using their shadows},
		\href{http://dx.doi.org/10.1140/epjc/s10052-020-8321-z}{Eur. Phys. J. C {\bfseries 80}, no.8, 717 (2020)},
		[\href{http://arxiv.org/abs/2005.00483}{arXiv:2005.00483 [gr-qc]}].
		
		\bibitem{Amarilla:2013sj}
		L.~Amarilla and E.~F.~Eiroa,
		{Shadow of a Kaluza-Klein rotating dilaton black hole},
		\href{http://dx.doi.org/10.1103/PhysRevD.87.044057}{Phys. Rev. D {\bfseries 87}, no.4, 044057 (2013)},
		[\href{http://arxiv.org/abs/1301.0532}{arXiv:1301.0532 [gr-qc]}].
		
		\bibitem{Atamurotov:2013sca}
		F.~Atamurotov, A.~Abdujabbarov and B.~Ahmedov,
		{Shadow of rotating non-Kerr black hole},
		\href{http://dx.doi.org/10.1103/PhysRevD.88.064004}{Phys. Rev. D {\bfseries 88}, no.6, 064004 (2013)}.
		
		\bibitem{Mishra:2019trb}
		A.~K.~Mishra, S.~Chakraborty and S.~Sarkar,
		{Understanding photon sphere and black hole shadow in dynamically evolving spacetimes},
		\href{http://dx.doi.org/10.1103/PhysRevD.99.104080}{Phys. Rev. D {\bfseries 99}, no.10, 104080 (2019)},
		[\href{http://arxiv.org/abs/1903.06376}{arXiv:1903.06376 [gr-qc]}].
		
		\bibitem{Papnoi:2014aaa}
		U.~Papnoi, F.~Atamurotov, S.~G.~Ghosh and B.~Ahmedov,
		{Shadow of five-dimensional rotating Myers-Perry black hole},
		\href{http://dx.doi.org/10.1103/PhysRevD.90.024073}{Phys. Rev. D {\bfseries 90}, no.2, 024073 (2014)},
		[\href{http://arxiv.org/abs/1407.0834}{arXiv:1407.0834 [gr-qc]}].
		
		\bibitem{Abdujabbarov:2015rqa}
		A.~Abdujabbarov, F.~Atamurotov, N.~Dadhich, B.~Ahmedov and Z.~Stuchl\'\i{}k,
		{Energetics and optical properties of 6-dimensional rotating black hole in pure Gauss\textendash{}Bonnet gravity},
		\href{http://dx.doi.org/10.1140/epjc/s10052-015-3604-5}{Eur. Phys. J. C {\bfseries 75}, no.8, 399 (2015)},
		[\href{http://arxiv.org/abs/1508.00331}{arXiv:1508.00331 [gr-qc]}].
		
		\bibitem{Kumar:2020owy}
		R.~Kumar and S.~G.~Ghosh,
		{Rotating black holes in $4D$ Einstein-Gauss-Bonnet gravity and its shadow},
		\href{http://dx.doi.org/10.1088/1475-7516/2020/07/053}{JCAP {\bfseries 07}, 053 (2020)},
		[\href{http://arxiv.org/abs/2003.08927}{arXiv:2003.08927 [gr-qc]}].
		
		\bibitem{Liu:2020ola}
		C.~Liu, T.~Zhu, Q.~Wu, K.~Jusufi, M.~Jamil, M.~Azreg-A\"\i{}nou and A.~Wang,
		{Shadow and quasinormal modes of a rotating loop quantum black hole},
		\href{http://dx.doi.org/10.1103/PhysRevD.101.084001}{Phys. Rev. D {\bfseries 101}, no.8, 084001 (2020)},
		[erratum: \href{http://dx.doi.org/10.1103/PhysRevD.103.089902}{Phys. Rev. D {\bfseries 103}, no.8, 089902 (2021)}],
		[\href{http://arxiv.org/abs/2003.00477}{arXiv:2003.00477 [gr-qc]}].
		
		\bibitem{Sarikulov:2022atq}
		F.~Sarikulov, F.~Atamurotov, A.~Abdujabbarov and B.~Ahmedov,
		{Shadow of the Kerr-like black hole},
		\href{http://dx.doi.org/10.1140/epjc/s10052-022-10711-4}{Eur. Phys. J. C {\bfseries 82}, no.9, 771 (2022)}.
		
		\bibitem{Ban:2024qsa}
		Z.~Ban, J.~Chen and J.~Yang,
		{Shadows of rotating black holes in effective quantum gravity},
		[\href{http://arxiv.org/abs/2411.09374}{arXiv:2411.09374 [gr-qc]}].
		
		\bibitem{Yang:2024nin}
		C.~Y.~Yang, M.~I.~Aslam, X.~X.~Zeng and R.~Saleem,
		{Shadow images of Ghosh-Kumar rotating black hole illuminated by spherical light sources and thin accretion disks},
		\href{http://dx.doi.org/10.1016/j.jheap.2025.01.017}{JHEAp {\bfseries 46}, 345 (2025)},
		[\href{http://arxiv.org/abs/2411.11807}{arXiv:2411.11807 [astro-ph.HE]}].
		
		\bibitem{Wang:2025ihg}
		X.~Wang, Z.~Zhao, X.~X.~Zeng and X.~Y.~Wang,
		{Revisiting the shadow of Johannsen-Psaltis black holes},
		\href{http://dx.doi.org/10.1103/PhysRevD.111.084054}{Phys. Rev. D {\bfseries 111}, no.8, 084054 (2025)},
		[\href{http://arxiv.org/abs/2501.08287}{arXiv:2501.08287 [gr-qc]}].
		
		\bibitem{Zeng:2025kqw}
		X.~X.~Zeng, C.~Y.~Yang, M.~I.~Aslam, R.~Saleem and S.~Aslam,
		{Kerr-like Black Hole Surrounded by Cold Dark Matter Halo: The Shadow Images and EHT Constraints},
		[\href{http://arxiv.org/abs/2505.07063}{arXiv:2505.07063 [gr-qc]}].
		
		\bibitem{Yunusov:2024xzu}
		O.~Yunusov, J.~Rayimbaev, F.~Sarikulov, M.~Zahid, A.~Abdujabbarov and Z.~Stuchl\'\i{}k,
		{Rotating charged black holes in EMS theory: shadow studies and constraints from EHT observations},
		\href{http://dx.doi.org/10.1140/epjc/s10052-024-13500-3}{Eur. Phys. J. C {\bfseries 84}, no.12, 1240 (2024)}.
		
		\bibitem{Zahid:2024hwi}
		M.~Zahid, O.~Yunusov, C.~Shen, J.~Rayimbaev and S.~Muminov,
		{Shadows and quasinormal modes of rotating black holes in Horndeski theory: Parameter constraints using EHT observations of M87* and Sgr A*},
		\href{http://dx.doi.org/10.1016/j.dark.2024.101734}{Phys. Dark Univ. {\bfseries 47}, 101734 (2025)}.
		
		\bibitem{Li:2024abk}
		X.~Q.~Li, H.~P.~Yan, X.~J.~Yue, S.~W.~Zhou and Q.~Xu,
		{Geodesic structure, shadow and optical appearance of black hole immersed in Chaplygin-like dark fluid},
		\href{http://dx.doi.org/10.1088/1475-7516/2024/05/048}{JCAP {\bfseries 05}, 048 (2024)},
		[\href{http://arxiv.org/abs/2401.18066}{arXiv:2401.18066 [gr-qc]}].
		
		\bibitem{Chen:2023wzv}
		S.~Chen and J.~Jing,
		{Kerr black hole shadows from axion-photon coupling},
		\href{http://dx.doi.org/10.1088/1475-7516/2024/05/023}{JCAP {\bfseries 05}, 023 (2024)},
		[\href{http://arxiv.org/abs/2310.06490}{arXiv:2310.06490 [gr-qc]}].
		
		\bibitem{Liu:2024lve}
		W.~Liu, D.~Wu and J.~Wang,
		{Shadow of slowly rotating Kalb-Ramond black holes},
		\href{http://dx.doi.org/10.1088/1475-7516/2025/05/017}{JCAP {\bfseries 05}, 017 (2025)}.
		[arXiv:2407.07416 [gr-qc]].
		
		\bibitem{Zahid:2025cfu}
		M.~Zahid, C.~Shen, J.~Rayimbaev, B.~Rahmatov, I.~Ibragimov, S.~Muminov and M.~Umaraliyev,
		{Rotating Yukawa-modified black holes: QNM and shadow studies},
		\href{http://dx.doi.org/10.1016/j.dark.2025.102124}{Phys.\ Dark Univ.\ {\bfseries 50}, 102124 (2025)}.
		
		\bibitem{Lambiase:2024lvo}
		G.~Lambiase, D.~J.~Gogoi, R.~C.~Pantig and A.~\"{O}vg\"{u}n,
		{Shadow and quasinormal modes of the rotating Einstein--Euler--Heisenberg black holes},
		\href{http://dx.doi.org/10.1016/j.dark.2025.101886}{Phys.\ Dark Univ.\ {\bfseries 48}, 101886 (2025)}.
		[arXiv:2406.18300 [gr-qc]].
		
		\bibitem{Ahmed:2025zdc}
		F.~Ahmed, S.~U.~Islam and S.~G.~Ghosh,
		{Shadows of rotating hairy black holes surrounded with quintessence and constraints from EHT observations},
		\href{http://dx.doi.org/10.1016/j.jheap.2025.100350}{JHEAp {\bfseries 46}, 100350 (2025)}.
		
		\bibitem{Zheng:2024ftk}
		H.~B.~Zheng, M.~Q.~Wu, G.~P.~Li and Q.~Q.~Jiang,
		{Shadows and accretion disk images of charged rotating black hole in modified gravity theory},
		\href{http://dx.doi.org/10.1140/epjc/s10052-025-13791-0}{Eur.\ Phys.\ J.\ C {\bfseries 85}, no.\ 1, 46 (2025)}.
		[arXiv:2411.10315 [gr-qc]].
		
		\bibitem{Siahaan:2025nlq}
		H.~M.~Siahaan,
		{Black hole shadows in accelerating Kerr--Newman--Taub--NUT and Braneworld spacetimes},
		\href{http://dx.doi.org/10.1140/epjc/s10052-025-14174-1}{Eur.\ Phys.\ J.\ C {\bfseries 85}, no.\ 4, 431 (2025)}.
		
		\bibitem{Newman:1965tw}
		E.~T.~Newman and A.~I.~Janis,
		\title{{Note on the Kerr Spinning Particle Metric}},
		\href{http://dx.doi.org/10.1063/1.1704350}{J. Math. Phys. \textbf{6}, 915 (1965)}.
		
		\bibitem{Ruffini:1971bza}
		R.~Ruffini and J.~A.~Wheeler,
		\href{http://dx.doi.org/10.1063/1.3022513}{Phys. Today \textbf{24}, no.1, 30 (1971)}.
		
		\bibitem{Bocharova:1970skc}
		N.~M.~Bocharova, K.~A.~Bronnikov and V.~N.~Melnikov, 
		{An exact solution of the system of Einstein and massless scalar field equations} (in Russian),
		Moscow Univ. Phys. Bull. \textbf{25}, 6 (1970).
		
		\bibitem{Bekenstein:1974sf}
		J.~D.~Bekenstein,
		\href{http://dx.doi.org/10.1016/0003-4916(74)90124-9}{Ann. Phys. \textbf{82}, 535-547 (1974)}.
		
		\bibitem{Martinez:2002ru}
		C.~Martínez, R.~Troncoso and J.~Zanelli,
		\href{http://dx.doi.org/10.1103/PhysRevD.67.024008}{Phys. Rev. D \textbf{67}, 024008 (2003)},
		[\href{http://arxiv.org/abs/hep-th/0205319}{arXiv:hep-th/0205319}].
		
		\bibitem{Martinez:2005di}
		C.~Martínez, J.~P.~Staforelli and R.~Troncoso,
		\href{http://dx.doi.org/10.1103/PhysRevD.74.044028}{Phys. Rev. D \textbf{74}, 044028 (2006)},
		[\href{http://arxiv.org/abs/hep-th/0512022}{arXiv:hep-th/0512022}].
		
		\bibitem{Astorino:2013sfa}
		M.~Astorino,
		\href{http://dx.doi.org/10.1103/PhysRevD.88.104027}{Phys. Rev. D \textbf{88}, no.10, 104027 (2013)},
		[\href{http://arxiv.org/abs/1307.4021}{arXiv:1307.4021 [gr-qc]}].
	
		\bibitem{NT}
		I.D. Novikov, K.S. Thorne, Astrophysics and black holes, in Les Houches Summer School of Theoretical Physics: Black Holes (1973), p. 343–550
		
		\bibitem{Collodel:2021gxu}
		L.~G.~Collodel, D.~D.~Doneva and S.~S.~Yazadjiev,
		{Circular Orbit Structure and Thin Accretion Disks around Kerr Black Holes with Scalar Hair},
		\href{http://dx.doi.org/10.3847/1538-4357/abe305}{Astrophys. J. {\bfseries 910}, no.1, 52 (2021)},
		[\href{http://arxiv.org/abs/2101.05073}{arXiv:2101.05073 [astro-ph.HE]}].
		
		\bibitem{Carter:1968rr}
		B.~Carter,
		{Global structure of the Kerr family of gravitational fields},
		\href{http://dx.doi.org/10.1103/PhysRev.174.1559}{Phys. Rev. {\bfseries 174}, 1559-1571 (1968)}.
		
		\bibitem{Hioki:2009na}
		K.~Hioki and K.~i.~Maeda,
		{Measurement of the Kerr Spin Parameter by Observation of a Compact Object's Shadow},
		\href{http://dx.doi.org/10.1103/PhysRevD.80.024042}{Phys. Rev. D {\bfseries 80}, 024042 (2009)},
		[\href{http://arxiv.org/abs/0904.3575}{arXiv:0904.3575 [astro-ph.HE]}].
		
	\end{thebibliography}
\end{document}